\documentclass[english,aps,prb,superscriptaddress,notitlepage]{revtex4-1}
\usepackage{lmodern}

\usepackage[T1]{fontenc}
\usepackage[latin9]{inputenc}
\usepackage{verbatim}
\usepackage{units}
\usepackage{mathtools}
\usepackage{amsmath}
\usepackage{amssymb}
\usepackage{wasysym}
\usepackage{graphicx}
\usepackage{esint}
\usepackage{babel}
\begin{document}

\title{Tunable magnetic phases in quasi-one-dimensional systems}

\author{Alfredo X. Sánchez}

\affiliation{Department of Physics and Beckman Institute, University of Illinois
at Urbana-Champaign, Urbana, Illinois 61801, USA}

\author{Jean-Pierre Leburton}

\email{jleburto@illinois.edu}

\affiliation{Department of Physics and Beckman Institute, University of Illinois
at Urbana-Champaign, Urbana, Illinois 61801, USA}

\affiliation{Department of Electrical and Computer Engineering, University of
Illinois at Urbana-Champaign, Urbana, Illinois 61801, USA}
\begin{abstract}
There has been considerable debate on the onset of exotic spin phenomena
in quantum wires due to enhanced many-body effects caused by the one-dimensional
(1D) alignment of charge carriers. We explain various observed spin
effects, such as a carrier density-dependent spin-flip in dilute quasi-1D
systems and the variability of the spin polarization in quantum point
contacts, by using an unrestricted Hartree-Fock approach with a three-dimensional
(3D) Coulomb interaction. The model dimensionality is critical in
identifying a complex pattern of magnetic phases varying with magnetic
field and confinement. In the limit of vanishing magnetic fields,
we show the emergence of a degenerate excited state with opposite
spin polarization above a confinement-dependent 1D concentration threshold,
which is consistent with observations of a conductance plateau at
half the conductance quantum $G_{0}/2=e^{2}/h$, even in the absence
of spin-orbit interactions.
\end{abstract}
\maketitle
Quantum wires (QWRs) are nanostructures characterized by two-dimensional
(2D) confinement exhibiting electronic modes transverse to the one-dimensional
(1D) motion of charge carriers. The existence of these transverse
modes has profound consequences on the interaction between carriers
and crystal dynamics, as well as amongst carriers themselves, which
uncovers a flurry of exciting properties. In the former case, carrier
scattering undergoes size effects\citep{Beenakker1991} that affect
the transport properties with important technological consequences
for device electronics\citep{Wolf2001}. In the latter case, the 1D
alignment of interacting particles enhances Hartree and exchange interactions
giving rise to exotic phenomena such as the formation of Wigner localization\citep{Matveev2004,Meyer2009}
or Luttinger liquids\citep{Tarucha1995,Yacoby1996}. During the last
two decades, a wide range of experiments stimulated by the observation
of a transport anomaly in semiconductor quantum point contacts (QPCs)\citep{Thomas1996,Nuttinck2000,Cronenwett2002}
and the demonstration of separate spin and charge excitations in QWRs\citep{Auslaender2005,Jompol2009}
have suggested the existence of spin-related transport or spontaneous
spin polarization in 1D systems\citep{Thomas1996,Kane1998,Thomas2000,Reilly2002,Starikov2003,Havu2004,Klironomos2006,Rokhinson2006,Berggren2008,Micolich2011,Lind2011},
as opposed to a Kondo-like effect due to the presence of a quasi-bound
state\citep{Cronenwett2002,Hirose2003,Rejec2006}. A spin-polarized
ground state, while forbidden in strictly 1D systems by the Lieb-Mattis
theorem\citep{Lieb1962a}, could be achieved in realistic QWRs and
QPCs, since these devices have two- and three-dimensional structures,
potentially giving rise to additional phenomena. This has spurred
interest in fully-electric manipulation of spin properties in quantum
wires and point contacts\citep{Debray2009,Wan2011}.

In this article, we show that 1D systems in longitudinal magnetic
fields and in the absence of spin-orbit interaction (as, for instance,
in a GaAs wire) can sustain a hierarchy of spin configurations depending
on carrier concentration, energy and confinement. Specifically, as
carrier concentration increases, the electron system in its ground
states evolves from a fully spin-polarized state to an unpolarized
state with a spin flip. In the limit of vanishing magnetic fields,
there exists a concentration threshold above which the electron system
exhibits an excited state with degenerate opposite-spin polarization.
We also show that this threshold varies as the strength of the confinement
is changed. The latter feature is particularly important for the technological
application of electrostatically-confined wires for which the spin
polarization can be controlled by electrical gating.

Let us consider a wire with its axis of symmetry along the $x$-axis.
Electrons are confined along the $y$- and $z$-directions (perpendicular
to the wire) by means of a potential $U_{\mathrm{conf}}\left(y,z\right)$,
which we model as a superposition of two parabolic wells, i.e. $U_{\mathrm{conf}}\left(y,z\right)=\frac{1}{2}m^{*}\omega_{y}^{2}y^{2}+\frac{1}{2}m^{*}\omega_{z}^{2}z^{2}$.
Here, $\omega_{y}$ and $\omega_{z}$ are the strengths of the confinement
along the $y$- and $z$-directions, respectively, and $m^{*}$ is
the electron effective mass. The wire is placed in a magnetic field
$\vec{B}=B_{0}\hat{x}$ parallel to the axis of the wire, with an
associated Zeeman term $U_{Z}=g\mu_{B}B_{x}\sigma$. ($g$ is the
effective electron $g$-factor, $\mu_{B}=\frac{q\hbar}{2m^{*}}$ is
the Bohr magneton in the wire, $q$ is the electron charge, and $\sigma$
is equal to $+1/2$ or $-1/2$ for spin-up or spin-down, respectively.)
For the sake of simplicity, the spin-orbit interaction is neglected.

The use of the unrestricted Hartree-Fock model in the extreme quantum
limit (i.e. when only one subband is populated) \citep{Sanchez2013}
results in the expression for the energy $E\left(k_{x},\sigma\right)$
of an electron in terms of its momentum $k_{x}$ and spin $\sigma$
(see Supplementary Methods), which reads:

\begin{equation}
E\left(k_{x},\sigma\right)=\frac{\hbar^{2}k_{x}^{2}}{2m^{*}}+\frac{1}{2}\hbar\omega_{y}+\frac{1}{2}\hbar\omega_{z}+\frac{1}{4}\left(\frac{\omega_{B}}{\omega_{y}}\right)\hbar\omega_{B}+g\mu_{B}B_{x}\sigma+U_{\mathrm{el}}\left[n_{0}\right]+U_{\mathrm{exch}}\left(k_{x},\sigma\right)\label{eq:E__kx_sigma}
\end{equation}

Here, $\omega_{B}=\frac{qB_{x}}{m^{*}}$ is the cyclotron frequency,
$U_{\mathrm{el}}$ is the Hartree term (which accounts for Coulomb
repulsion amongst electrons and is proportional to the total concentration
$n_{0}$) and $U_{\mathrm{exch}}$ is the exchange term. $U_{\mathrm{el}}$
and $U_{\mathrm{exch}}$ depend on the overlap function $\zeta_{ab}\left(p\right)$
which, in turn, is determined by the strength of the lateral confinement
and the shape of the electron wavefunction. (See Supplementary Methods.)
The pair of two integral equations for $E\left(k_{x},\uparrow\right)$
and $E\left(k_{x},\downarrow\right)$ described by Eq. (\ref{eq:E__kx_sigma})
is then solved to yield the spin-dependent concentrations $n_{\sigma}$
($n_{\uparrow}$, $n_{\downarrow}$) for a fixed total electron concentration
$n_{0}=n_{\uparrow}+n_{\downarrow}$.

At zero temperature, both $n_{\uparrow}$ and $n_{\downarrow}$ are
associated with a (positive) spin-dependent Fermi wavevector $k_{f\left(\sigma\right)}=\pi n_{\sigma}$,
so that the Fermi energy $E_{f}=E\left(k_{f\left(\sigma\right)}\right)=E_{f}\left(n_{\sigma}\right)$.
Then, after setting $E_{f}\left(n_{\uparrow}\right)=E_{f}\left(n_{\downarrow}\right)$,
the relation between $n_{\uparrow}$ and $n_{\downarrow}$ can be
written in terms of $n_{0}$ and $\Delta n_{\sigma}\equiv n_{\sigma}-n_{0}/2$,
i.e.

\begin{equation}
\frac{\pi^{2}}{2}n_{0}\Delta n_{\sigma}+\sigma\frac{qgB_{x}}{2\hbar}-\frac{q^{2}m^{*}}{64\pi^{2}\epsilon\hbar^{2}}\int_{2\pi\left(n_{0}/2-\Delta n_{\sigma}\right)}^{2\pi\left(n_{0}/2+\Delta n_{\sigma}\right)}dp\,\zeta_{ab}\left(p\right)=0\label{eq:n0_Dns}
\end{equation}

This equation is discussed in more detail in the Supplementary Methods
and holds only for $0\leq\left|\Delta n_{\sigma}\right|\leq n_{0}/2$,
since $0\leq n_{\sigma}\leq n_{0}$. If Eq. (\ref{eq:n0_Dns}) yields
the solution $\Delta n_{\uparrow\left(\downarrow\right)}>n_{0}/2$,
the identity $k_{f\left(\downarrow\left(\uparrow\right)\right)}=\pi n_{\downarrow\left(\uparrow\right)}$
is no longer valid because $n_{\downarrow\left(\uparrow\right)}$
would then be negative, and thus must be set to zero in order to still
satisfy Eq. (\ref{eq:E__kx_sigma}). In this case, $n_{\downarrow\left(\uparrow\right)}=n_{0}$,
which corresponds to full spin polarization.

\begin{figure}[h]
\noindent \begin{centering}
\includegraphics[width=3in]{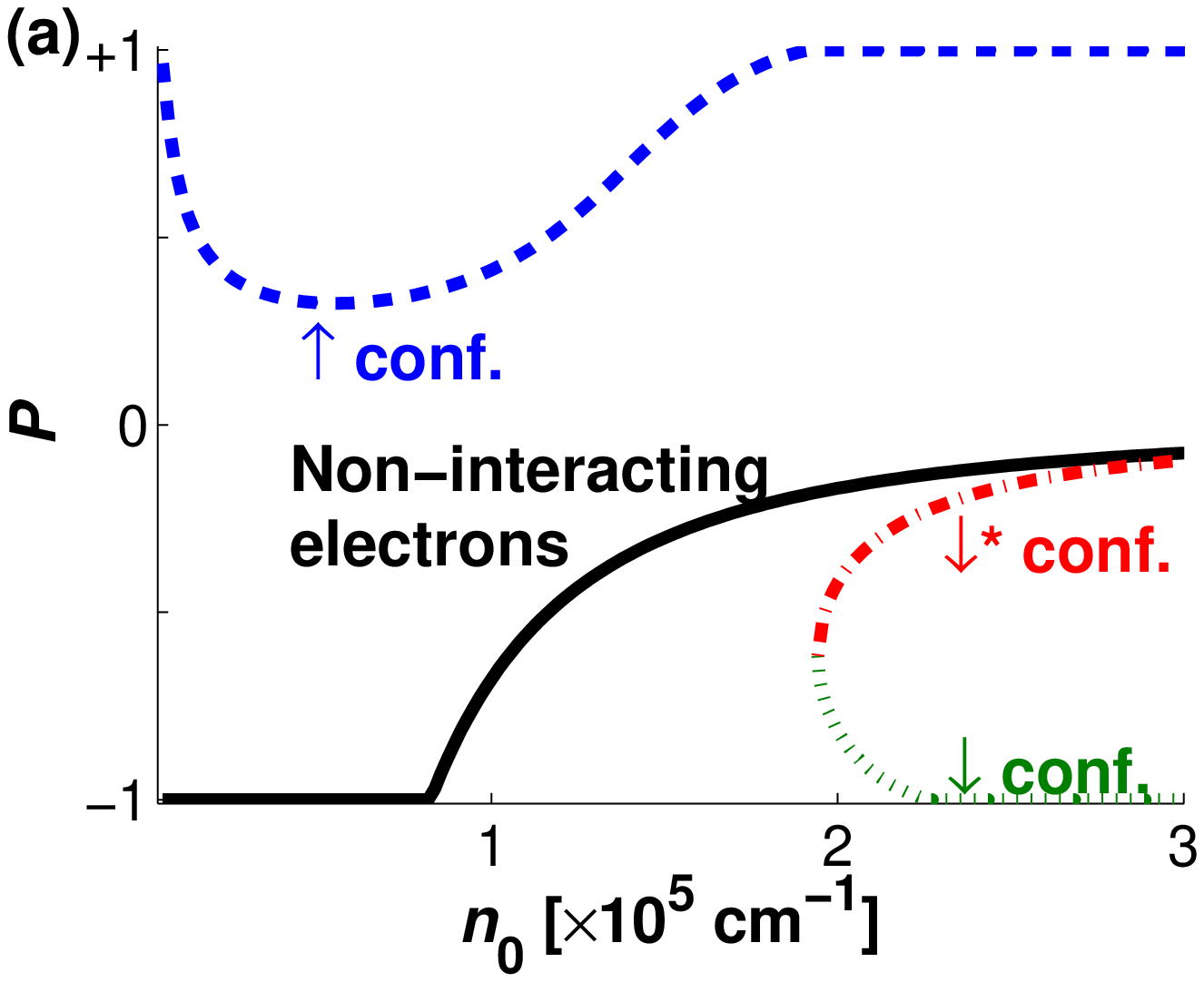}\includegraphics[width=3in]{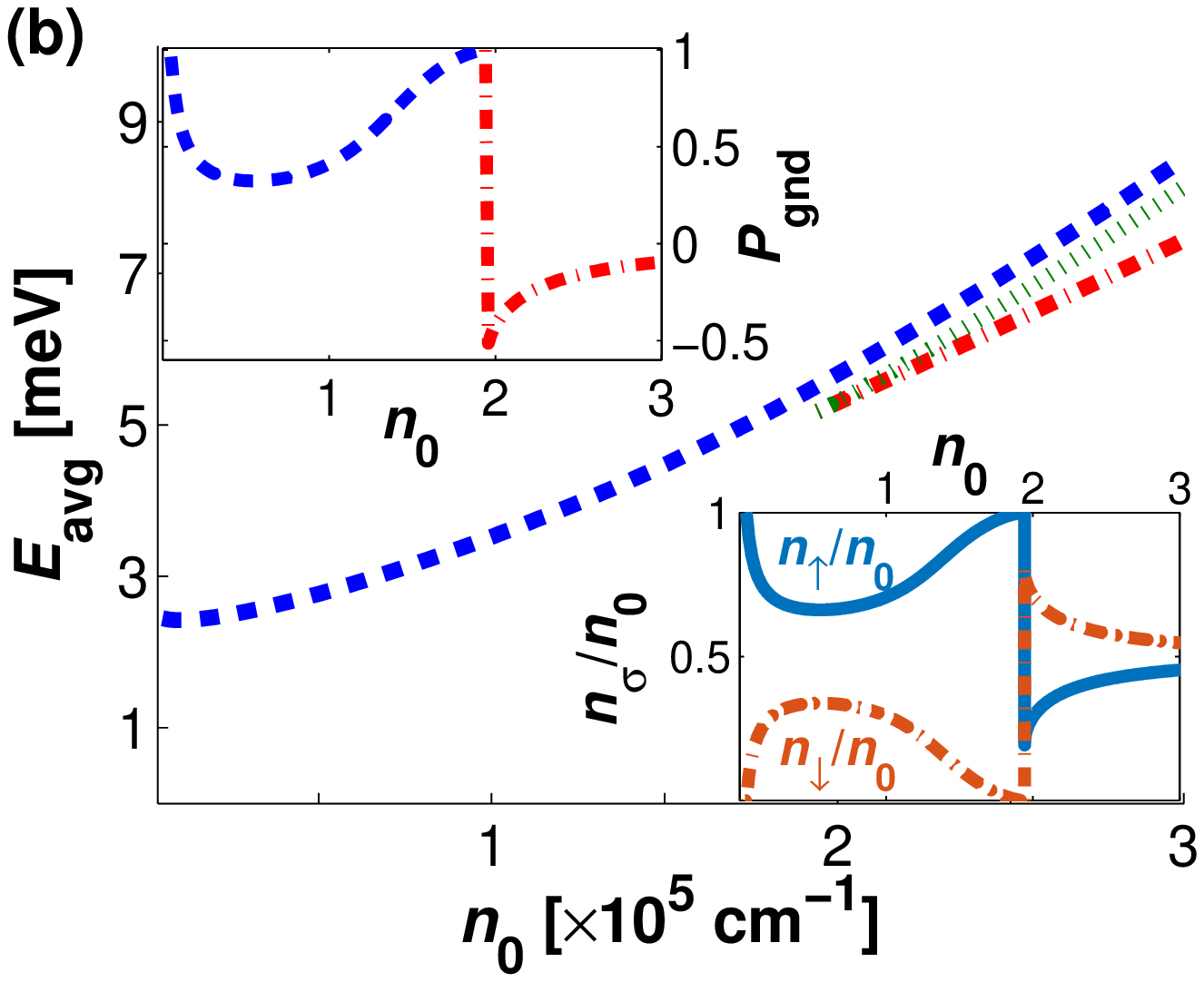}
\par\end{centering}

\protect\caption{\label{fig:P__vs__n0}\textbf{Spin-polarized regimes.} (a) Polarization
$P=\left(n_{\uparrow}-n_{\downarrow}\right)/n_{0}$ as a function
of the total concentration $n_{0}$ at $B_{x}=\unit[1]{T}$ and $T=\unit[0]{K}$
for non-interacting electrons (solid line) and the three possible
polarization configurations for interacting electrons: $\uparrow$
(dashed), $\downarrow$ (dotted) and $\downarrow^{*}$ (dash-dotted). }
\end{figure}

In order to illustrate our model, we consider a GaAs quantum wire
with $\left|g\right|=0.44$, $m^{*}=0.067m_{0}$ and $\epsilon=12.9\epsilon_{0}$\citep{Hess2000},
and where the transverse confinement strength is set to $\hbar\omega_{y}=\hbar\omega_{z}=\unit[2.0]{meV}$.
Figure \ref{fig:P__vs__n0}(a) shows the polarization $P=\left(n_{\uparrow}-n_{\downarrow}\right)/n_{0}$
for interacting and non-interacting electrons, as a function of the
total concentration $n_{0}$, at $B_{x}=\unit[1]{T}$ and zero temperature.
The solid line in Figure \ref{fig:P__vs__n0}(a) corresponds to the
spin polarization for non-interacting electrons, i.e. $\zeta_{ab}\left(p\right)=0$
in Eq. (\ref{eq:n0_Dns}). In this case $P$ is negative because of
the Zeeman interaction, $U_{Z}=g\mu_{B}B_{x}\sigma$, which lowers
the potential energy of spin-down electrons, and hence $n_{\downarrow}$
exceeds $n_{\uparrow}$ at the same Fermi energy. For concentrations
$n_{0}\leq n_{B}=\unit[0.817\times10^{5}]{cm^{-1}}$, there is complete
spin-down polarization and $P=-1$. For $n_{0}$ exceeding $n_{B}$,
one gets $n_{\sigma}=\frac{n_{0}}{2}\left[1-\left(\frac{n_{B}}{n_{0}}\right)^{2}\mathrm{sign}\left(gB_{x}\sigma\right)\right]$
(as described in the Supplementary Methods), so $P=-\left(\frac{n_{B}}{n_{0}}\right)^{2}$
and $-1<P<0$, which corresponds to partial spin-down polarization.
As $n_{0}$ increases to infinity, $P$ slowly approaches zero. This
is due to the fact that the kinetic energy term $T_{x\left(\sigma\right)}=\frac{\hbar^{2}k_{f\left(\sigma\right)}^{2}}{2m^{*}}=\frac{\hbar^{2}\pi^{2}n_{\sigma}^{2}}{2m^{*}}$,
which increases with $n_{\sigma}$, contributes predominantly to the
total energy at high concentration since $U_{Z}$ is independent of
$n_{0}$. As a result, the Zeeman splitting induced by $U_{Z}$ becomes
insignificant.

The dashed line in Figure \ref{fig:P__vs__n0}(a) is one of the solutions
to Eq. (\ref{eq:E__kx_sigma}) for interacting electrons, which we
call the \textquotedblleft $\uparrow$'' or ``up'' configuration,
as $P>0$. For this ``$\uparrow$'' configuration, the wire is fully
spin-polarized when $n_{0}\leq n_{0}^{\,\mathrm{full(min)}}=\unit[4.51\times10^{3}]{cm^{-1}}$
and $n_{0}\geq n_{0}^{\,\mathrm{full(max)}}=\unit[1.92\times10^{5}]{cm^{-1}}$.
These limiting values are obtained by solving Eq. (\ref{eq:n0_Dns}),
setting $\Delta n_{\uparrow}=+n_{0}/2$. Between those two concentrations,
there is partial spin polarization, with a minimum $P=0.32$ at $n_{0}=\unit[0.54\times10^{5}]{cm^{-1}}$.
Spin polarization is opposite to the non-interacting case because
of the existence of the exchange interaction, which lowers the zero-point
energy of the 1D energy subband. At very low $n_{0}$, the exchange
energy $U_{\mathrm{exch}}$ dominates the kinetic energy $T_{x}$,
but since $E_{f}\left(n_{\uparrow}\right)=E_{f}\left(n_{\downarrow}\right)$,
$n_{\uparrow}$ increases and $n_{\downarrow}$ decreases (otherwise,
$U_{\mathrm{exch}\left(\downarrow\right)}$ would be much more negative
than $U_{\mathrm{exch}\left(\uparrow\right)}$, which would cause
$E_{f}\left(n_{\downarrow}\right)$ to drop far below $E_{f}\left(n_{\uparrow}\right)$).

\begin{figure}[h]
\noindent \centering{}\includegraphics[width=3in]{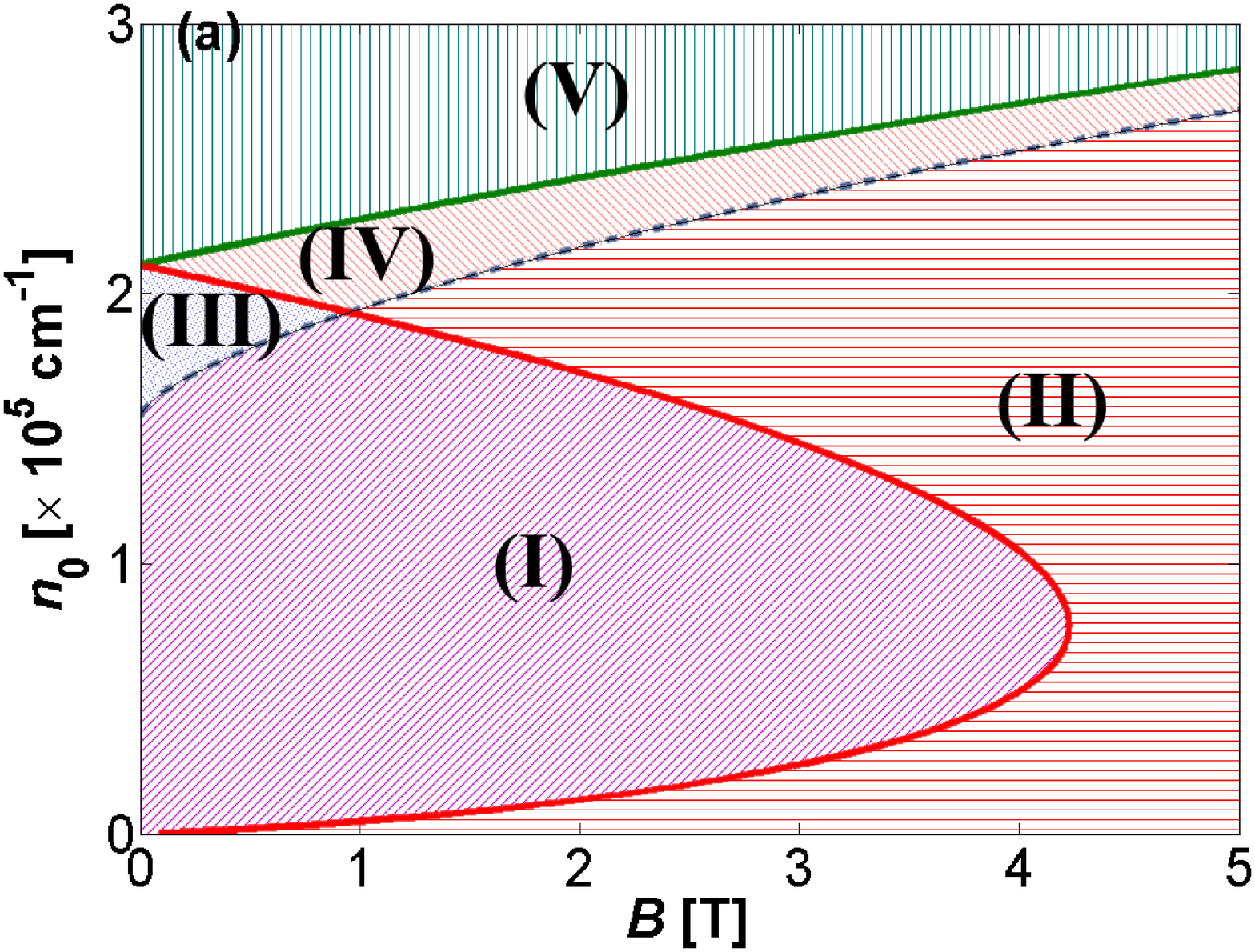}\includegraphics[width=3in]{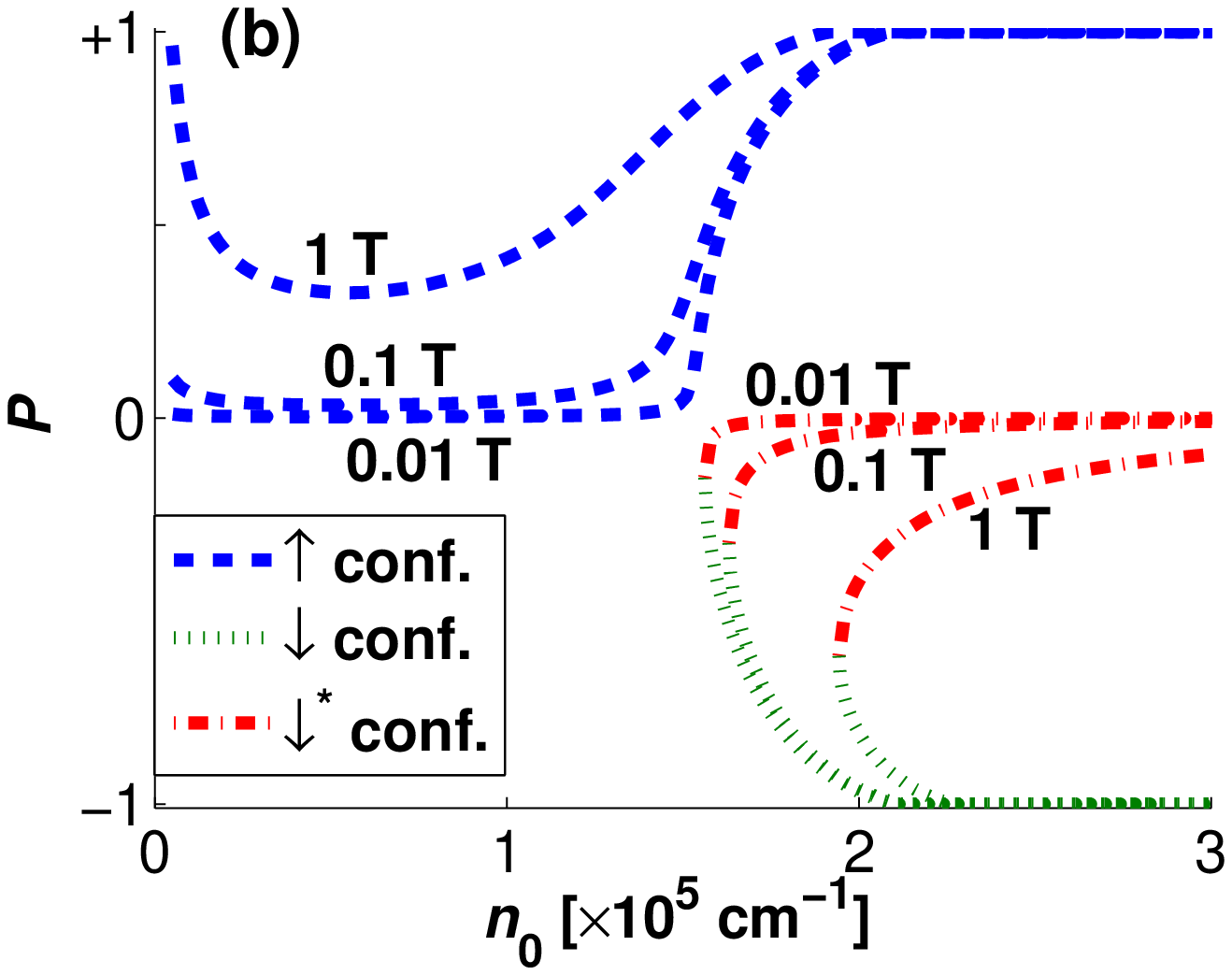}\protect\caption{\label{fig:n0__v__Bx}\textbf{Magnetic field dependence of the spin-polarized
regimes.} (a) Magnetic phases diagram of the distinct spin-polarized
regimes as a function of $n_{0}$ and $B_{x}$: (I) $\uparrow$ configuration
only, partially polarized; (II) $\uparrow$ configuration only, fully
polarized; (III) $\uparrow$ (partial), $\downarrow$ (partial), $\downarrow^{*}$
(partial); (IV) $\uparrow$ (full), $\downarrow$ (partial), $\downarrow^{*}$
(partial); (V) $\uparrow$ (full), $\downarrow$ (full), $\downarrow^{*}$
(partial). (b) Polarization vs. $n_{0}$ for different magnetic field
strengths. In all cases, $\hbar\omega_{y}=\hbar\omega_{z}=\unit[2.0]{meV}$
and $T=\unit[0]{K}$.}
\end{figure}

\begin{figure}[h]
\noindent \begin{centering}
\includegraphics[width=3in]{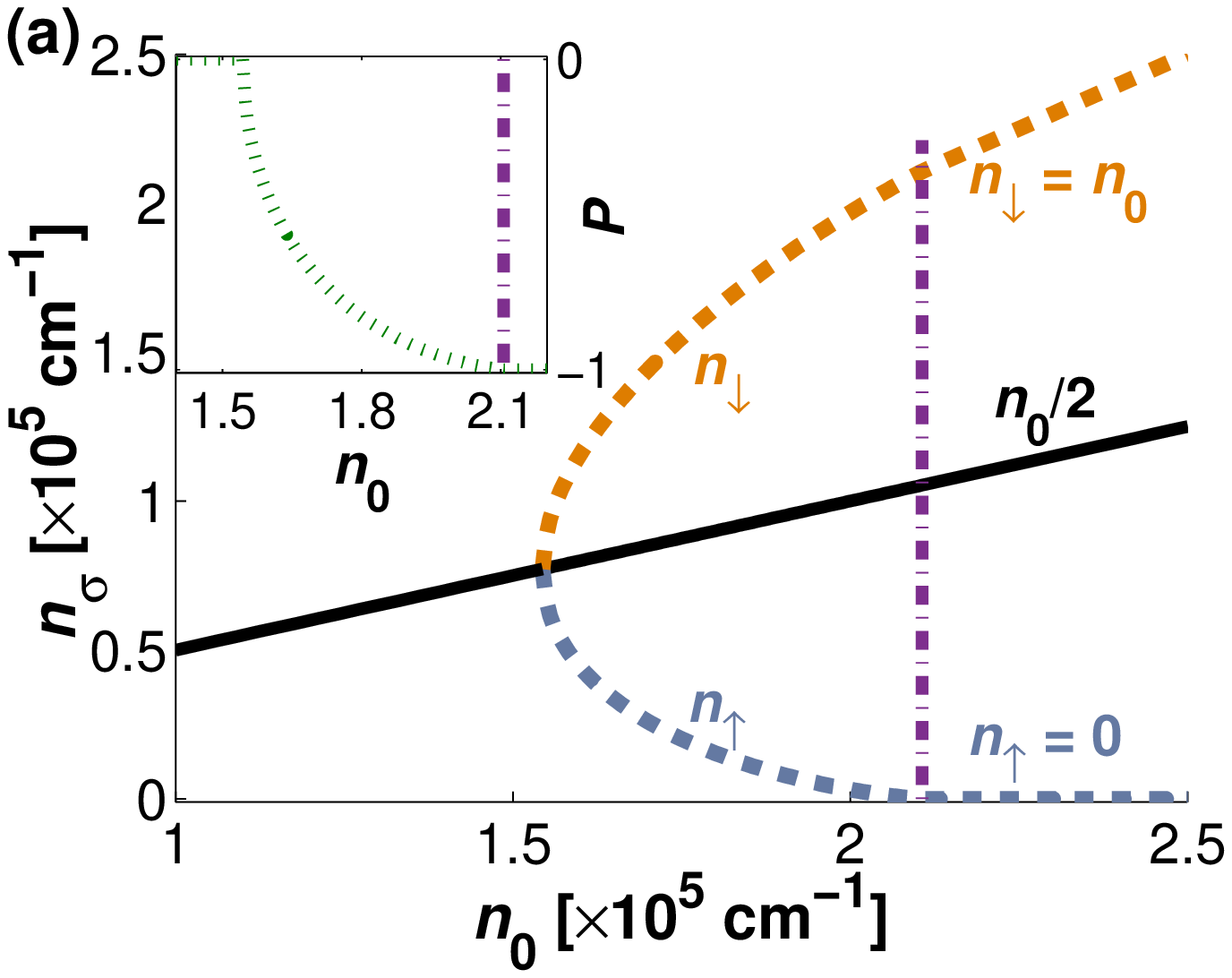}\includegraphics[width=3in]{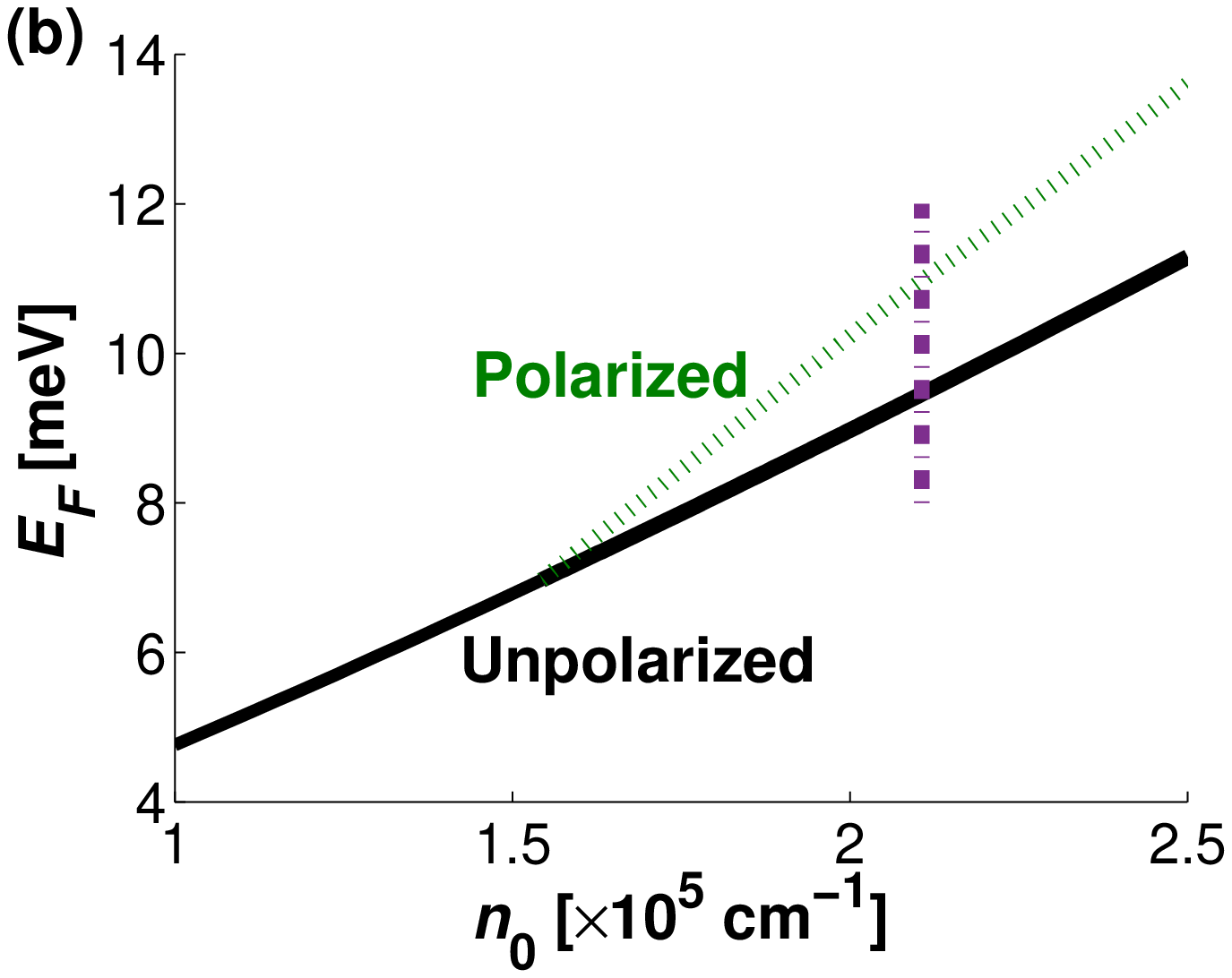}
\par\end{centering}

\protect\caption{\label{fig:ns__v__n0__B0}\textbf{Spin-polarized regimes at $B_{x}=0$.}
(a) Spin-polarized concentrations $n_{\uparrow}$ and $n_{\downarrow}$
vs. $n_{0}$ for the $\downarrow$ configuration when $B_{x}=0$.
Inset: polarization of the $\downarrow$ configuration vs. $n_{0}$.
(b) Fermi energy vs. $n_{0}$ for both the polarized (solid) and unpolarized
(dotted) regimes. The vertical dash-dotted line marks the concentration
for full polarization ($n_{0}^{\,\mathrm{full}}=\unit[2.1\times10^{5}]{cm^{-1}}$).
$\hbar\omega_{y}=\hbar\omega_{z}=\unit[2.0]{meV}$ and $T=\unit[0]{K}$.}
\end{figure}

\begin{figure}[h]
\noindent \begin{centering}
\includegraphics[width=3in]{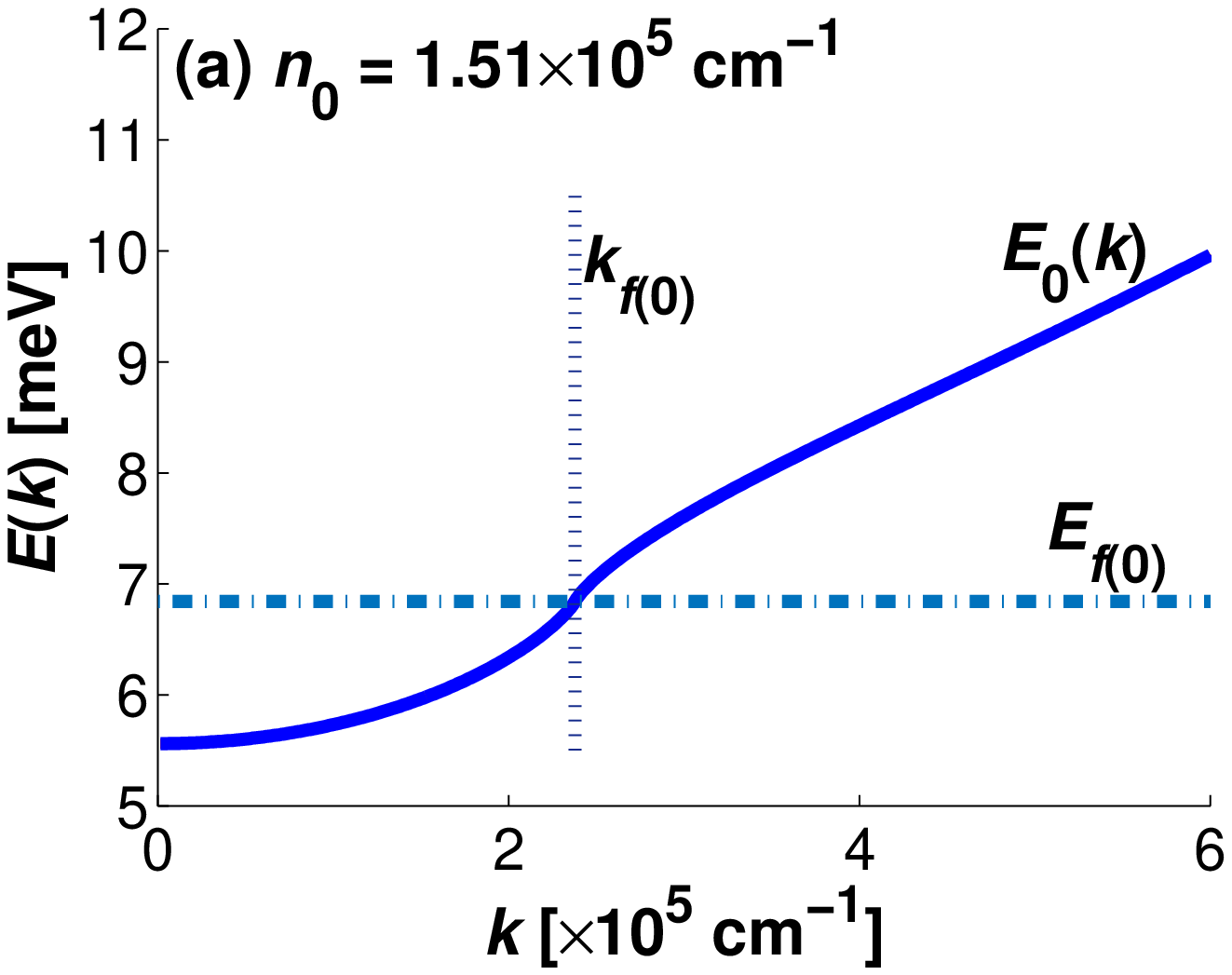}\includegraphics[width=3in]{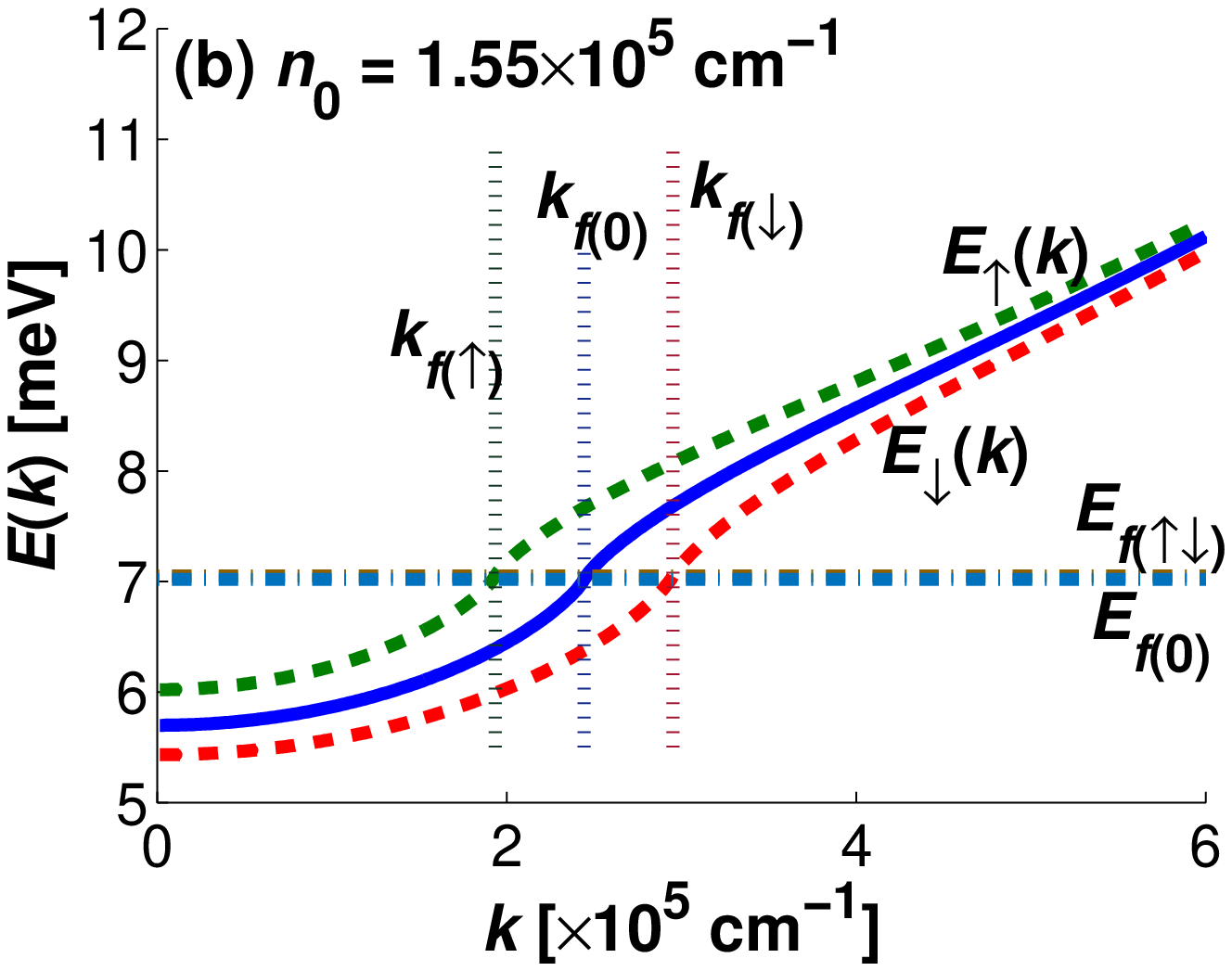}
\par\end{centering}

\noindent \begin{centering}
\includegraphics[width=3in]{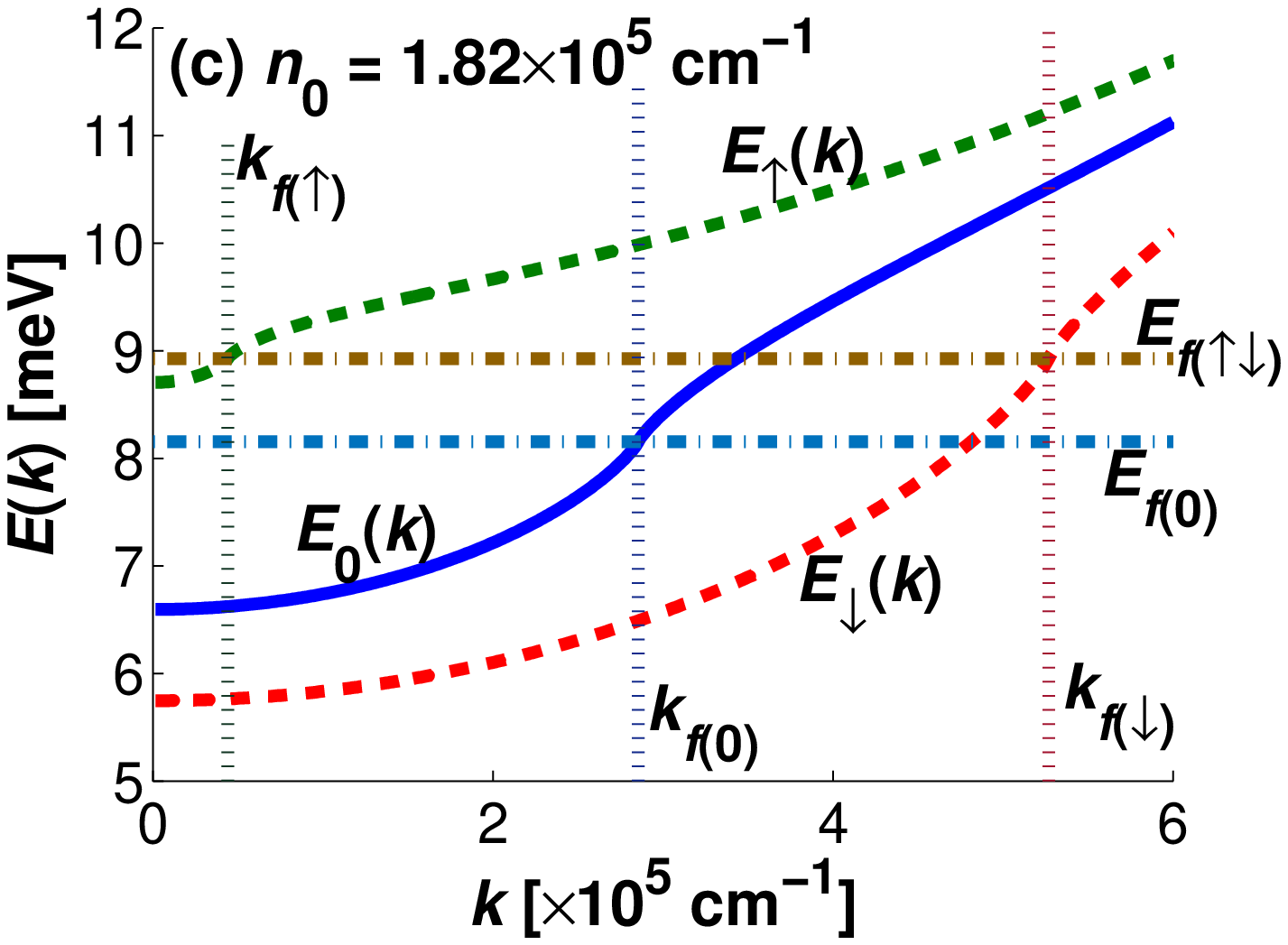}
\par\end{centering}

\protect\caption{\label{fig:E__v__k}\textbf{Dispersion relation.} Energy vs. wavevector
when $B_{x}=0$, for three different concentrations: (a) just below
the threshold for polarization; (b) just above the threshold; (c)
significantly above the threshold. $\hbar\omega_{y}=\hbar\omega_{z}=\unit[2.0]{meV}$
and $T=\unit[0]{K}$. $E_{0}\left(k\right)$, $E_{f\left(0\right)}$
and $k_{f\left(0\right)}$ are, respectively, the energy and the Fermi
wavevector in the unpolarized configuration, while $E_{\sigma}\left(k\right)$
and $k_{f\left(\sigma\right)}$are the energy and the Fermi wavevector
in the polarized configuration for electrons of spin $\sigma\left(=\uparrow,\downarrow\right)$.}
\end{figure}

\begin{figure}[h]
\noindent \begin{centering}
\includegraphics[width=3in]{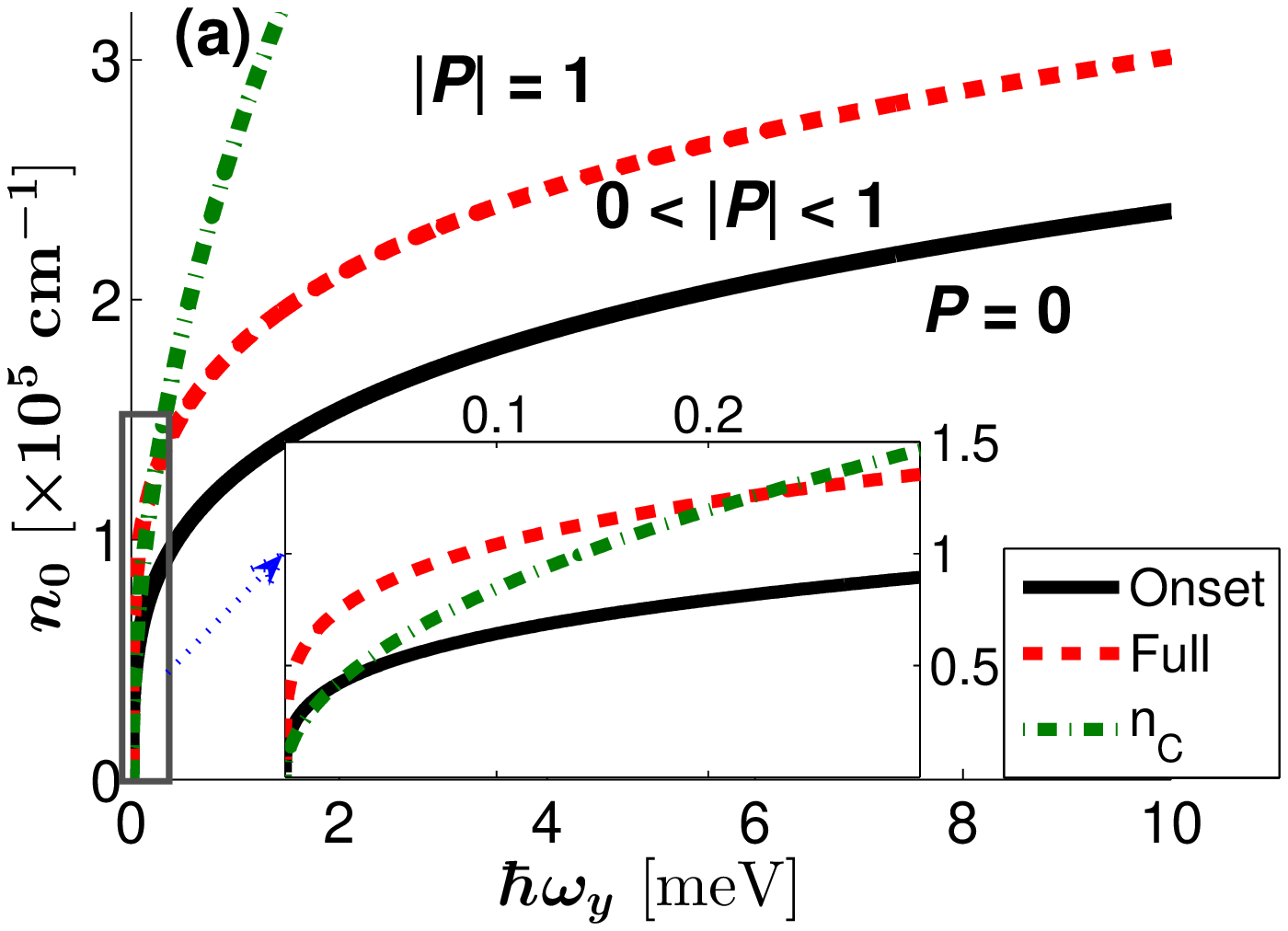}\includegraphics[width=3in]{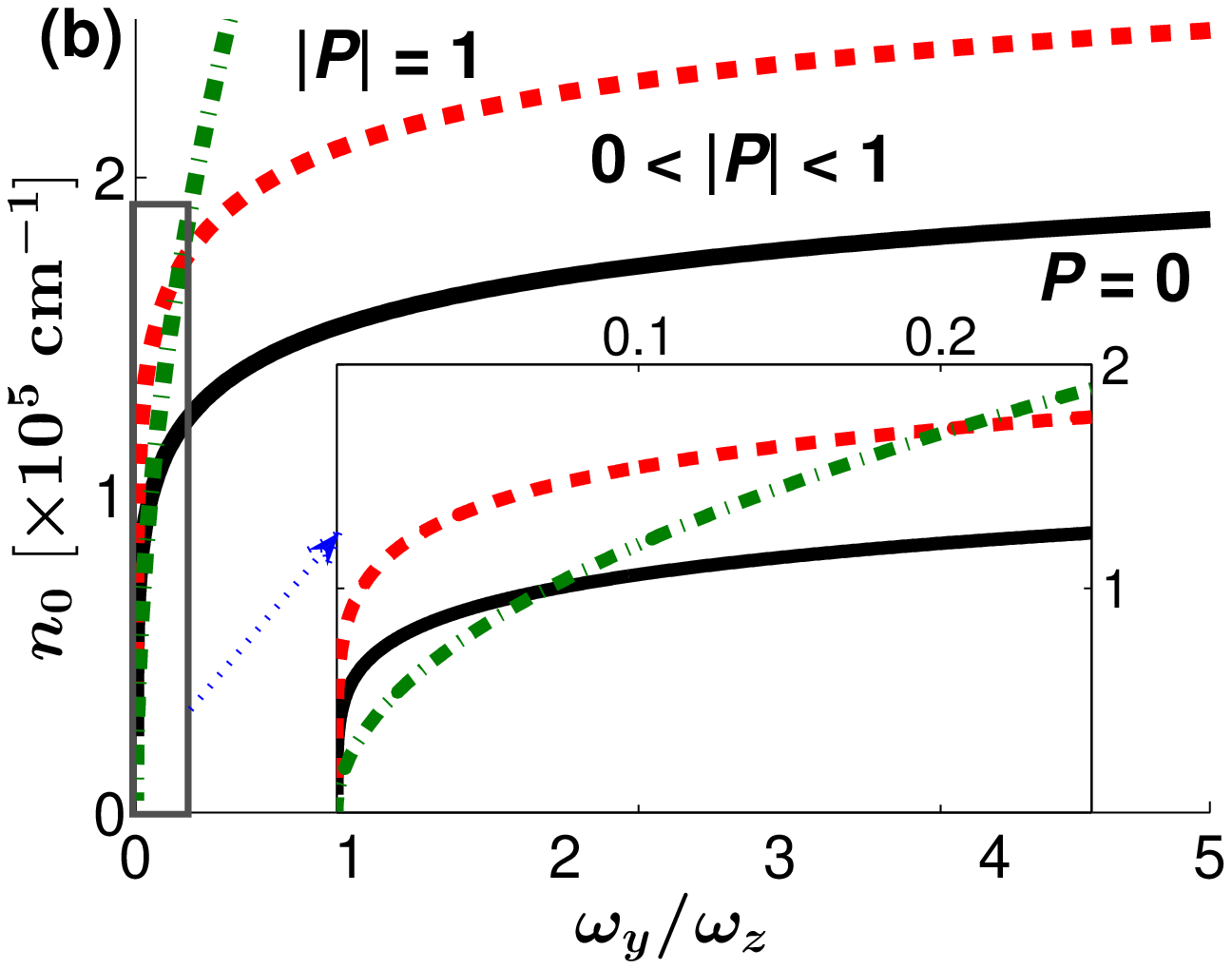}
\par\end{centering}

\protect\caption{\label{fig:n0_onset_full__v__conf}\textbf{Confinement dependence
of the spin-polarized regimes.} Concentration $n_{0}$ at the onset
of polarization (solid line) and when full polarization is achieved
(dashed line) as a function of: (a) $\hbar\omega_{y}$ for the case
$\hbar\omega_{y}=\hbar\omega_{z}\equiv\hbar\omega$; (b) $\omega_{y}/\omega_{z}$,
with $\hbar\omega_{z}=\unit[2]{meV}$. $B_{x}=0$ and $T=0$ in both
cases. Insets: close-ups of regions $\hbar\omega<\unit[0.3]{meV}$
(in (a)) and $\omega_{y}/\omega_{z}<0.25$ (in (b)) showing the maximum
concentration $n_{C}$ for which only one subband is occupied (dash-dotted
line).}
\end{figure}

For $n_{0}\geq n_{0}^{\,\mathrm{onset}}\left(=\unit[1.94\times10^{5}]{cm^{-1}}\right)$,
two additional configurations emerge, both with $P<0$: ``$\downarrow$''
or ``down'' (dotted line in Figure \ref{fig:P__vs__n0}(a)) and
\textquotedblleft $\downarrow^{*}$\textquotedblright{} or \textquotedblleft down-star\textquotedblright{}
(dash-dotted line). In both cases, $P=-0.62$ at the threshold concentration
$n_{0}^{\,\mathrm{onset}}$, but the two configurations differ as
$n_{0}$ increases. In the ``$\downarrow$'' regime, the polarization
becomes stronger (i.e. more negative) until the wire is fully polarized
($P=-1$) above $n_{0}^{\,\mathrm{full}}=\unit[2.27\times10^{5}]{cm^{-1}}$
(a value also obtained from Eq. (\ref{eq:n0_Dns}) with $\Delta n_{\downarrow}=+n_{0}/2$).
Meanwhile, in the ``$\downarrow^{*}$'' regime, spin polarization
is weakened and tends to zero (approaching the Zeeman splitting for
non-interacting electrons) as $n_{0}$ goes to infinity. Here we point
out that the specific numerical values of the concentrations $n_{0}^{\,\mathrm{onset}}$,
$n_{0}^{\,\mathrm{full}}$, etc., as well as the energies displayed
on figures \ref{fig:P__vs__n0}-\ref{fig:E__v__k}, are all confinement-dependent
and will change for different confinement strengths (see Fig. \ref{fig:n0_onset_full__v__conf}).

The different high-concentration behaviors of the three configurations
($P=1$, $P=-1$ and $P\rightarrow0$ for the $\uparrow$, $\downarrow$
and $\downarrow^{*}$ configurations, respectively) result from the
interplay between the kinetic energy $T_{x}$, the exchange interaction
$U_{\mathrm{exch}}$ and the Zeeman splitting $U_{Z}$ to satisfy
the condition $E_{f}\left(n_{\uparrow}\right)=E_{f}\left(n_{\downarrow}\right)$.
At high $n_{0}$, $U_{\mathrm{exch}}$ dominates $U_{Z}$, which can
be attenuated in three ways: (1) If $n_{\downarrow}$ is smaller than
$n_{\uparrow}$, which prevents the combined contribution $U_{\mathrm{exch}\left(\downarrow\right)}+U_{Z\left(\downarrow\right)}$
from becoming too negative; this leads to full spin-up polarization
($\uparrow$ configuration). (2) If $n_{\downarrow}$ is instead larger
than $n_{\uparrow}$, which reduces the kinetic energy term $T_{x\left(\uparrow\right)}$
relative to $T_{x\left(\downarrow\right)}$, thus lowering $E_{f}\left(n_{\uparrow}\right)$
and resulting in full spin-down polarization ($\downarrow$ configuration).
(3) If $n_{\downarrow}$ is almost equal to $n_{\uparrow}$ (i.e.
$P\rightarrow0$, just like for non-interacting electrons) in order
to balance $T_{x\left(\downarrow\right)}$ and $T_{x\left(\uparrow\right)}$,
since $T_{x}$ dominates $U_{\mathrm{exch}}$(and $U_{Z}$) at high
concentrations; this is the behavior of the $\downarrow^{*}$ configuration.

Whether the electrons in the ground state of the system are in the
$\uparrow$, $\downarrow$ or $\downarrow^{*}$ configuration depends
on which of these three configurations has the lowest energy. Figure
\ref{fig:P__vs__n0}(b) shows the average energy per electron as a
function of $n_{0}$ for all three possibilities. If $n_{0}$ is above
$n_{0}^{\,\mathrm{onset}}$, when all three solutions are possible,
the $\downarrow^{*}$ configuration has the lowest energy, with an
energy difference with the other two solutions that grows as $n_{0}$
increases. While the $\downarrow$ and $\downarrow^{*}$ configurations
both emerge with the same energy $E_{\mathrm{avg}}=\unit[5.2]{meV}$
at $n_{0}^{\,\mathrm{onset}}$, the separation between these two grows
up to $\unit[0.7]{meV}$ at $n_{0}=\unit[3\times10^{5}]{cm^{-1}}$.
The $\uparrow$ configuration has the highest energy, exceeding that
of the $\downarrow$ solution by a constant value of $\unit[0.4]{meV}$.
However, below the $n_{0}^{\,\mathrm{onset}}$ threshold, $\downarrow^{*}$
is forbidden, so the system assumes the $\uparrow$ configuration.
As a consequence, at $n_{0}^{\,\mathrm{onset}}$, the ground state
changes abruptly from a positive to a negative polarization. This
spin-flip process is shown on the insets of figure \ref{fig:P__vs__n0}(b).
The polarization $P_{\mathrm{gnd}}$ of the ground state (top-left
inset) drops suddenly from $+1$ ($\uparrow$ configuration, dashed
line) to $-0.6$ ($\downarrow^{*}$, dash-dotted line) at $n_{0}^{\,\mathrm{onset}}$.
The bottom-right inset displays the ratio $n_{\sigma}/n_{0}$ for
the ground state of the system. Below $n_{0}^{\,\mathrm{onset}}$,
the system is in the $\uparrow$-configuration where $n_{\uparrow}$
(solid line) exceeds $n_{\downarrow}$ (dashed line). In this region,
there is full spin-up polarization for concentrations below $\unit[4.51\times10^{3}]{cm^{-1}}$
and above $\unit[1.92\times10^{5}]{cm^{-1}}$. Between these two values
$n_{\uparrow}$ reaches a minimum at $n_{0}=\unit[0.54\times10^{5}]{cm^{-1}}$
where only two-thirds of the electrons in the wire have spin-up. At
$n_{0}^{\,\mathrm{onset}}$, the system switches to the $\downarrow^{*}$
configuration, for which $n_{\downarrow}>n_{\uparrow}$. At this spin-reversal
point, three-fourths of the electrons have spin down, but as $n_{0}$
increases the difference between $n_{\uparrow}$ and $n_{\downarrow}$
decreases.

Figure \ref{fig:n0__v__Bx}(a) displays the various spin polarization
regimes in the wire as both the magnetic fields and the concentrations
are varied. Only the $\uparrow$ configuration is present in regions
(I) and (II), with either partial or full polarization, respectively.
These two regions are separated by a solid, almost parabolically-shaped
line. For $B_{x}>\unit[4.22]{T}$ the $\uparrow$ configuration is
completely polarized regardless of the concentration. The black dashed
line corresponds to the minimum concentration $n_{0}^{\,\mathrm{onset}}$
for the emergence of the $\downarrow$ and $\downarrow^{*}$ configurations.
In regions (III) and (IV) the $\uparrow$ configuration is, respectively,
partially and fully polarized. Finally, region (V) lies above the
topmost solid line that indicates the minimum concentration $n_{0}^{\,\mathrm{full}}$
for which the $\downarrow$ configuration is fully polarized. In this
region only the $\downarrow^{*}$ configuration is partially polarized.

The presence of different polarization regimes in Fig. \ref{fig:n0__v__Bx}
is consistent with direct measurements of the spin polarization in
quantum point contacts\citep{Rokhinson2006} showing their variability
as the magnetic field and concentration are changed. Furthermore,
the sudden polarization reversal predicted in our model (Fig. \ref{fig:P__vs__n0}(b))
also explains the abrupt rearrangement of the spin-up and spin-down
levels under a strong in-plane magnetic field as observed by Graham
et al \citep{Graham2007} and previously interpreted as an exchange-driven
magnetic phase transition \citep{Berggren2005,Graham2007,Lassl2007}.
Indeed, at $n_{0}^{\,\mathrm{onset}}$, the polarization of the ground
state changes from positive to negative, so the energy of spin-up
electrons suddenly rises above that of spin-down electrons, leading
to the observed depopulation of the spin-up subband.

Figure \ref{fig:n0__v__Bx}(b) shows the effect of vanishing magnetic
fields on the different spin polarizations in the wire, which evolve
from a complex of configurations ($\uparrow$, $\downarrow$, and
$\downarrow^{*}$) for $B_{x}\neq0$ to a situation that sustains
an unpolarized state co-existing with two degenerate and symmetrically
spin-polarized regimes at high concentrations for $B_{x}=0$. One
can clearly see that the $\uparrow$ spin polarization at low concentration,
and the $\downarrow^{*}$ spin polarization at high concentration
collapse to an unpolarized configuration, while the $\downarrow$
and $\uparrow$ configurations are symmetric relative to each other.
At the same time, the concentration threshold $n_{0}^{\,\mathrm{onset}}$
decreases to a lower common value for up- and down-spin polarization.
As will be seen in Fig. \ref{fig:ns__v__n0__B0}(b), these two configurations
are degenerate, characterized by a single Fermi level and equally
probable, which is consistent with the absence of net magnetic moment
in the wire.

The existence of spin polarization degeneracy when $B_{x}=0$ can
be shown from Eq. (\ref{eq:n0_Dns}) where, asides from the trivial
solution $\Delta n_{\sigma}=0$ that corresponds to a spin-unpolarized
electron density, one obtains the approximate solution for a non-zero
$\Delta n_{\sigma}$ in the limit $\left|\Delta n_{\sigma}\right|\ll n_{0}$
(see Supplementary Methods):
\begin{equation}
\Delta n_{\sigma}\approx\pm\sqrt{\frac{3}{\zeta_{ab}^{\prime\prime}\left(\pi n_{0}\right)}\left[a^{*}n_{0}-\frac{\zeta_{ab}\left(\pi n_{0}\right)}{2\pi^{2}}\right]}\label{eq:Dns__v__n0__B0__approx}
\end{equation}

Here, $a^{*}=\frac{4\pi\epsilon\hbar^{2}}{q^{2}m^{*}}$ is the effective
Bohr radius and $\zeta_{ab}^{\prime\prime}\left(\pi n_{0}\right)=\left[\frac{d^{2}\zeta_{ab}\left(p\right)}{dp^{2}}\right]_{p=\pi n_{0}}$.
This solution is only valid for $n_{0}\geq n_{0\left(B_{x}=0\right)}^{\,\mathrm{onset}}$,
where $n_{0\left(B_{x}=0\right)}^{\,\mathrm{onset}}$ satisfies the
identity obtained by setting $\Delta n_{\sigma}=0$ in Eq. (\ref{eq:Dns__v__n0__B0__approx}):
\begin{equation}
\frac{\zeta_{ab}\left(\pi n_{0\left(B_{x}=0\right)}^{\,\mathrm{onset}}\right)}{\pi n_{0\left(B_{x}=0\right)}^{\,\mathrm{onset}}}=2\pi a^{*}\label{eq:n0onset_B0_zeta_astar}
\end{equation}

Figure \ref{fig:ns__v__n0__B0}(a) is a plot of the spin-polarized
concentrations in the wire for the $\downarrow$ configuration ($n_{\downarrow}>n_{\uparrow}$)
when $B_{x}=0$. The solid line corresponds to the unpolarized case
($n_{\downarrow}=n_{\uparrow}=n_{0}/2$). The spin-polarized regime,
represented by the dashed curves, emerges at $n_{0}=n_{0}^{\,\mathrm{onset}}\left(=\unit[1.54\times10^{5}]{cm^{-1}}\right)$.
Above this threshold $n_{\downarrow}$ continues to increase with
$n_{0}$ until the wire becomes fully polarized ($n_{\downarrow}=n_{0}$,
$n_{\uparrow}=0$) for $n_{0}\geq n_{0}^{\,\mathrm{full}}\left(=\unit[2.1\times10^{5}]{cm^{-1}}\right)$.
The inset shows the polarization corresponding to this configuration
as it changes from $0$ at $n_{0}=n_{0}^{\,\mathrm{onset}}$ to $-1$
at $n_{0}=n_{0}^{\,\mathrm{full}}$.

Both spin-polarized configurations ($\uparrow$ and $\downarrow$)
have the same energy, which is generally higher than that of the unpolarized
regime. This is shown in Fig. \ref{fig:ns__v__n0__B0}(b), which is
a plot of the Fermi energy as a function of total concentration. At
$n_{0}^{\,\mathrm{onset}}$ the Fermi energy is $\unit[6.97]{meV}$
for all configurations, but as $n_{0}$ increases the energy of the
polarized configurations rises more rapidly, so at $n_{0}=n_{0}^{\,\mathrm{full}}$
the difference in energies between the two cases is about $\unit[1.5]{meV}$.
Therefore, at $T=\unit[0]{K}$, the ground state of the electrons
in the wire will be unpolarized, in agreement with the Lieb-Mattis
theorem\citep{Lieb1962a}. However, at higher temperatures the system
can be excited to one of the spin-polarized regimes (which are indistinguishable
from each other in the absence of a $B$-field), especially for concentrations
close to $n_{0}^{\,\mathrm{onset}}$ when the energy difference between
the polarized and unpolarized configurations is lower.

The energy-momentum relation for spin-up and spin-down electrons at
$B_{x}=0$ is illustrated in Figures \ref{fig:E__v__k}(a-c) for three
different concentrations in the $\downarrow$ configuration ($P<0$).
In all cases there is an inflection point at the Fermi wave vector
caused by the exchange interaction, which indicates the presence of
a maximum in the carrier velocity, a minimum in the 1D density of
states, and an infinite effective mass. In Fig. \ref{fig:E__v__k}(a),
for which $n_{0}<n_{0}^{\,\mathrm{onset}}\left(=\unit[1.54\times10^{5}]{cm^{-1}}\right)$,
there is a single energy-momentum curve that, once the spin-polarized
configuration emerges for $n_{0}>n_{0}^{\,\mathrm{onset}}$, splits
into three curves, one of them above and another one below the unpolarized
dispersion relation. In Fig. \ref{fig:E__v__k}(b) the concentration
is just above the threshold, so that both polarized concentrations
exist below their Fermi level, which sits slightly higher than its
unpolarized counterpart. In Fig. \ref{fig:E__v__k}(c), the curve-splitting
is more significant. The topmost dispersion relation, corresponding
to spin-up electrons, is almost entirely above the spin-polarized
Fermi level, which, in turn, is much higher than the unpolarized level.
In both Fig. \ref{fig:E__v__k}(b) and \ref{fig:E__v__k}(c), the
spin polarization tapers off at high wavevectors/energies, indicating
that the effect weakens with carrier energy and concentration.

The emergence of density-dependent spin-polarized configurations is
in agreement with previous experimental observations of a small conductance
plateau around $0.5G_{0}=e^{2}/h$ that appears in quantum wires for
specific electron concentrations\citep{Kane1998,Thomas2000,Reilly2002}.
According to our model, this plateau arises from the gap between the
energies $E\left(k,\sigma\right)$ of spin-up and spin-down electrons
in the polarized configurations, as shown in Fig. \ref{fig:E__v__k}.
Indeed, when the concentration is increased past the polarization
threshold, e.g. by changing a gate voltage in a GaAs/AlGaAs heterostructure,
one of the spin channels will gradually close, leading to a narrow
step at $e^{2}/h$ instead of $2e^{2}/h$. The predicted spin-polarized
regime is an excited state, though, and the system is expected to
fall back to the unpolarized configuration (which has a lower average
energy; see Fig. \ref{fig:ns__v__n0__B0}(b)) as the density is increased,
especially at low temperatures. The fact that the observed $e^{2}/h$
feature is narrower than the $2e^{2}/h$ plateau and is enhanced with
increasing temperature is consistent with previous conductance measurements
\citep{Thomas2000,Reilly2002}.

The theoretical results also support the interpretation of a spin-polarized
excited state as the origin of the anomalous conductance structure
at 0.7 G0 observed in quantum point contacts (QPCs) \citep{Thomas1996,Thomas2000,Rokhinson2006}.
Since the spin-polarized regimes occur in the high-energy excited
states, the QPC potential barrier for these states is higher than
that of the unpolarized configuration, reducing the conductance for
those spin-polarized electrons. Furthermore, the electron concentration
in spin-polarized configurations can be lowered by changing the gate
voltage in the QPC heterostructure, inducing a further reduction of
the conductance. These two facts are consistent with the temperature-enhanced
pinning of the conductance near $0.7G_{0}$ at specific gate voltages.

As mentioned before, the concentrations at the onset of polarization
($n_{0}^{\,\mathrm{onset}}$) and at full polarization ($n_{0}^{\,\mathrm{full}}$)
also depend on the strength of the transverse confinement along the
$y$- and $z$-directions, i.e. on the size of the cross-section of
the wire. This is entirely due to the exchange interaction, since
Eq. (\ref{eq:n0_Dns}) predicts a confinement-independent threshold
$n_{0}^{\,\mathrm{onset}}$ for non-interacting electrons ($\zeta_{ab}\left(p\right)\rightarrow0$).
Figure \ref{fig:n0_onset_full__v__conf}(a) displays the values of
$n_{0}^{\,\mathrm{onset}}$ (solid line) and $n_{0}^{\,\mathrm{full}}$
(dashed line) as a function of the confinement strength $\hbar\omega_{y}$
for a wire with a circular cross section ($\hbar\omega_{y}=\hbar\omega_{z}\equiv\hbar\omega$)
and at $B_{x}=0$. If $\hbar\omega$ increases, both $n_{0}^{\,\mathrm{onset}}$
and $n_{0}^{\,\mathrm{full}}$ increase, although the rate of increase
slows down and becomes almost constant when $\hbar\omega>\unit[2]{meV}$.
Furthermore, the range of $n_{0}$ for which partial polarization
exists ($0<\left|P\right|<1$, corresponding to region (III) of Fig.
\ref{fig:n0__v__Bx}(a)) becomes wider when the confinement is stronger.
These data indicate that the wider the wire, i.e. the weaker the confinement,
the lower the concentration for which polarization is possible, provided
that the system remains in the extreme quantum limit. To satisfy this
condition, the concentration must be smaller than a critical value
$n_{C}=\frac{2}{\pi\hbar}\sqrt{2m^{*}\left(\hbar\omega_{y}\right)}$,
since $\hbar\omega_{y}$ is the energy above the bottom of the first
subband at which the second subband will start to become populated
(when $\omega_{y}\leq\omega_{z}$). The inset shows a close-up of
the $n_{0}^{\,\mathrm{onset}}$ and $n_{0}^{\,\mathrm{full}}$ curves
for small values of $\hbar\omega$, as well as a plot of $n_{C}$
(dash-dotted line). For $\hbar\omega\apprle\unit[0.02]{meV}$, $n_{0}^{\,\mathrm{onset}}>n_{C}$
and the system is outside the scope of the model.

The effects of asymmetric confinement on $n_{0}^{\,\mathrm{onset}}$
and $n_{0}^{\,\mathrm{full}}$ are portrayed on Figure \ref{fig:n0_onset_full__v__conf}(b)
as a function of the confinement strength ratio $\omega_{y}/\omega_{z}$
for $\hbar\omega_{z}=\unit[2]{meV}$, in the absence of a magnetic
field. Both $n_{0}^{\,\mathrm{onset}}$ and $n_{0}^{\,\mathrm{full}}$
increase with the ratio, very rapidly for $\omega_{y}<\omega_{z}$
and at a much slower rate for $\omega_{y}>\omega_{z}$. This reflects
the fact that the overlap function $\zeta_{ab}$ grows with stronger
confinement (see Supplementary Figure \ref{fig:S1}). For instance,
for $\hbar\omega_{y}=\hbar\omega_{z}=\unit[10]{meV}$, $n_{0}^{\,\mathrm{onset}}=\unit[2.34\times10^{5}]{cm^{-1}}$,
whereas for $\hbar\omega_{y}=\unit[10]{meV}$ and $\hbar\omega_{z}=\unit[2]{meV}$
(corresponding to $\omega_{y}/\omega_{z}=5$), $n_{0}^{\,\mathrm{onset}}=\unit[1.87\times10^{5}]{cm^{-1}}$.
As in Figure \ref{fig:n0_onset_full__v__conf}(a), the spin-polarized
configurations are accessible until $n_{0}^{\,\mathrm{onset}}>n_{C}$;
as shown in the inset of Figure \ref{fig:n0_onset_full__v__conf}(b),
this happens when $\omega_{y}/\omega_{z}\apprle0.068$ or $\hbar\omega_{y}\apprle\unit[0.14]{meV}$.

Our model shows that electron-electron interactions in quasi-1D systems
with longitudinal magnetic fields lead to complex interweaved spin-polarized
regimes which differ from previous works\citep{Lind2011}, e.g. magnetic
configurations opposite to those expected from pure Zeeman splitting
in low-concentration quantum wires. These new results are the consequence
of considering an exact 3D $1/r$-Coulomb potential as opposed to
a Dirac point potential\citep{Lassl2007,Lind2011}, which enables
us to include the interaction between electrons with parallel spins
(prohibited with a Dirac interaction\citep{Lind2011}) in the energy-momentum
relation, Eq. (\ref{eq:E__kx_sigma}). Specifically, at zero $B$-field,
our theory shows that two symmetric and degenerate spin-polarized
configurations exist for all values of carrier concentration (and
Fermi energy) above a concentration threshold, even in the absence
of spin-orbit interaction. The existence of the spin-polarized excited
state in quasi-1D systems, more readily accessible as temperature
is increased, is consistent with the emergence of a narrow conductance
step near $0.5\times2e^{2}/h$ at low temperatures ($\sim\unit[50-100]{mK}$)\citep{Kane1998,Thomas2000,Reilly2002},
and its subsequent evolution into a broader feature near $0.7\times2e^{2}/h$
as the temperature is increased to a few kelvins in both quantum wires
and quantum point contacts\citep{Thomas1996,Thomas2000,Kristensen2000,Reilly2002}.
The collective spin states described in our model also account for
the observations of separate spin modes in coupled quantum wires that
were originally attributed to Luttinger-liquid behavior\citep{Auslaender2005,Jompol2009}.
Our findings, however, result from the nonlinear nature of the energy
dispersion relation as the particle concentration increases, thereby
outlining the limitations of the low-energy, linear-dispersion Luttinger
model\citep{Imambekov2012}. Such limitations are revealed through
the observation of fast energy relaxation of particles in quantum
wires, a process forbidden by Luttinger-liquid theory\citep{Barak2010}.
Moreover, our results do not conform to the Wigner-crystal picture\citep{Matveev2004,Meyer2009}
either, since it precludes the presence of spin-polarized states in
the absence of a magnetic field. Indeed, in this picture, the condition
for polarization is that the Zeeman energy $\left|U_{Z}\right|=g\mu_{B}B/2$
exceed the exchange energy $\left|U_{\mathrm{exch}}\right|$\citep{Deshpande2010},
in contrast to our findings.

Furthermore, our theory highlights the importance of a full 3D approach
to account for the sensitivity of the spin-polarized configurations
to varying confinement strength and asymmetry. In this context, our
model uses a pair of parabolic potentials for transverse confinement,
which is suitable for elliptical wire cross-sections. It is, however,
general and valid for other geometries, for which the overlap function
$\zeta_{ab}\left(p\right)$, critical in setting the concentration
threshold $n_{0}^{\,\mathrm{onset}}$ for the different spin-polarized
regimes (Eq. (\ref{eq:n0onset_B0_zeta_astar})), would need to be
evaluated with the corresponding wave functions. This is particularly
relevant for spintronics applications, as it enables the design of
quantum wires with specific spin-polarization characteristics by changing
the shape and confinement of the 1D channel. For instance, in a GaAs/AlGaAs
heterostructure, the desired range of polarizations can be set by
choosing a suitable physical separation between the split gates and
a specific acceptor density in GaAs. Subsequent fine-tuning of the
polarization could be achieved by simply adjusting the split-gate
potential bias.

Interestingly, our prediction that spin polarization is possible in
symmetric quantum wires contrasts with the findings of Debray et al
\citep{Debray2009}, whose experiments observed a spin-polarized current
in a QPC only when the confinement potential was strongly asymmetric.
It should, however, be pointed out that their QPC device is made with
InAs, a material characterized by a strong intrinsic spin-orbit interaction.
For this reason, the authors of that work attribute the emergence
of spin polarization to a lateral spin-orbit coupling (which is a
function of both the electron momentum and the confinement potential),
as opposed to the Rashba spin-orbit interaction (which only depends
on momentum). The extension of our model to incorporate both types
of spin-orbit couplings is a topic of future investigation. However,
the main objective of this work was to show that, even in the absence
of such couplings, a variety of magnetic phases is achieved in quasi-1D
systems by electric manipulation alone.

\section*{Methods}

We calculate the 1D energy-momentum relation within the unrestricted
Hartree-Fock approximation, considering only the extreme quantum limit
(i.e. when only the lowest-energy subband of the confinement potential
is occupied), as described in the Supplementary Methods. All numerical
calculations (e.g. solving for $\Delta n_{\sigma}$ in Eq. (\ref{eq:n0_Dns}))
were done using MATLAB.
\begin{acknowledgments}
This work is supported by a grant from the Research Board at the University
of Illinois at Urbana-Champaign (UIUC). A.X.S. also thanks the Department
of Physics at UIUC for their continued support during his studies.
\end{acknowledgments}
\bibliographystyle{apsrev4-1}
\bibliography{C:/Documents/UIUC/Research/References/AllReferences}

\begin{thebibliography}{35}%
\makeatletter
\providecommand \@ifxundefined [1]{%
 \@ifx{#1\undefined}
}%
\providecommand \@ifnum [1]{%
 \ifnum #1\expandafter \@firstoftwo
 \else \expandafter \@secondoftwo
 \fi
}%
\providecommand \@ifx [1]{%
 \ifx #1\expandafter \@firstoftwo
 \else \expandafter \@secondoftwo
 \fi
}%
\providecommand \natexlab [1]{#1}%
\providecommand \enquote  [1]{``#1''}%
\providecommand \bibnamefont  [1]{#1}%
\providecommand \bibfnamefont [1]{#1}%
\providecommand \citenamefont [1]{#1}%
\providecommand \href@noop [0]{\@secondoftwo}%
\providecommand \href [0]{\begingroup \@sanitize@url \@href}%
\providecommand \@href[1]{\@@startlink{#1}\@@href}%
\providecommand \@@href[1]{\endgroup#1\@@endlink}%
\providecommand \@sanitize@url [0]{\catcode `\\12\catcode `\$12\catcode
  `\&12\catcode `\#12\catcode `\^12\catcode `\_12\catcode `\%12\relax}%
\providecommand \@@startlink[1]{}%
\providecommand \@@endlink[0]{}%
\providecommand \url  [0]{\begingroup\@sanitize@url \@url }%
\providecommand \@url [1]{\endgroup\@href {#1}{\urlprefix }}%
\providecommand \urlprefix  [0]{URL }%
\providecommand \Eprint [0]{\href }%
\providecommand \doibase [0]{http://dx.doi.org/}%
\providecommand \selectlanguage [0]{\@gobble}%
\providecommand \bibinfo  [0]{\@secondoftwo}%
\providecommand \bibfield  [0]{\@secondoftwo}%
\providecommand \translation [1]{[#1]}%
\providecommand \BibitemOpen [0]{}%
\providecommand \bibitemStop [0]{}%
\providecommand \bibitemNoStop [0]{.\EOS\space}%
\providecommand \EOS [0]{\spacefactor3000\relax}%
\providecommand \BibitemShut  [1]{\csname bibitem#1\endcsname}%
\let\auto@bib@innerbib\@empty
%</preamble>
\bibitem [{\citenamefont {Beenakker}\ and\ \citenamefont {van
  Houten}(1991)}]{Beenakker1991}%
  \BibitemOpen
  \bibfield  {author} {\bibinfo {author} {\bibfnamefont {C.}~\bibnamefont
  {Beenakker}}\ and\ \bibinfo {author} {\bibfnamefont {H.}~\bibnamefont {van
  Houten}},\ }in\ \href {\doibase DOI: 10.1016/S0081-1947(08)60091-0} {\emph
  {\bibinfo {booktitle} {Semiconductor Heterostructures and Nanostructures}}},\
  \bibinfo {series} {Solid State Physics}, Vol.~\bibinfo {volume} {44},\
  \bibinfo {editor} {edited by\ \bibinfo {editor} {\bibfnamefont
  {H.}~\bibnamefont {Ehrenreich}}\ and\ \bibinfo {editor} {\bibfnamefont
  {D.}~\bibnamefont {Turnbull}}}\ (\bibinfo  {publisher} {Academic Press},\
  \bibinfo {address} {Eindhoven},\ \bibinfo {year} {1991})\ pp.\ \bibinfo
  {pages} {1 -- 228}\BibitemShut {NoStop}%
\bibitem [{\citenamefont {Wolf}\ \emph {et~al.}(2001)\citenamefont {Wolf},
  \citenamefont {Awschalom}, \citenamefont {Buhrman}, \citenamefont {Daughton},
  \citenamefont {von Molnár}, \citenamefont {Roukes}, \citenamefont
  {Chtchelkanova},\ and\ \citenamefont {Treger}}]{Wolf2001}%
  \BibitemOpen
  \bibfield  {author} {\bibinfo {author} {\bibfnamefont {S.~A.}\ \bibnamefont
  {Wolf}}, \bibinfo {author} {\bibfnamefont {D.~D.}\ \bibnamefont {Awschalom}},
  \bibinfo {author} {\bibfnamefont {R.~A.}\ \bibnamefont {Buhrman}}, \bibinfo
  {author} {\bibfnamefont {J.~M.}\ \bibnamefont {Daughton}}, \bibinfo {author}
  {\bibfnamefont {S.}~\bibnamefont {von Molnár}}, \bibinfo {author}
  {\bibfnamefont {M.~L.}\ \bibnamefont {Roukes}}, \bibinfo {author}
  {\bibfnamefont {A.~Y.}\ \bibnamefont {Chtchelkanova}}, \ and\ \bibinfo
  {author} {\bibfnamefont {D.~M.}\ \bibnamefont {Treger}},\ }\href {\doibase
  10.1126/science.1065389} {\bibfield  {journal} {\bibinfo  {journal}
  {Science}\ }\textbf {\bibinfo {volume} {294}},\ \bibinfo {pages} {1488}
  (\bibinfo {year} {2001})}\BibitemShut {NoStop}%
\bibitem [{\citenamefont {Matveev}(2004)}]{Matveev2004}%
  \BibitemOpen
  \bibfield  {author} {\bibinfo {author} {\bibfnamefont {K.~A.}\ \bibnamefont
  {Matveev}},\ }\href {\doibase 10.1103/PhysRevLett.92.106801} {\bibfield
  {journal} {\bibinfo  {journal} {Phys. Rev. Lett.}\ }\textbf {\bibinfo
  {volume} {92}},\ \bibinfo {pages} {106801} (\bibinfo {year}
  {2004})}\BibitemShut {NoStop}%
\bibitem [{\citenamefont {Meyer}\ and\ \citenamefont
  {Matveev}(2009)}]{Meyer2009}%
  \BibitemOpen
  \bibfield  {author} {\bibinfo {author} {\bibfnamefont {J.~S.}\ \bibnamefont
  {Meyer}}\ and\ \bibinfo {author} {\bibfnamefont {K.~A.}\ \bibnamefont
  {Matveev}},\ }\href {http://stacks.iop.org/0953-8984/21/i=2/a=023203}
  {\bibfield  {journal} {\bibinfo  {journal} {J. Phys.: Condens. Matter}\
  }\textbf {\bibinfo {volume} {21}},\ \bibinfo {pages} {023203} (\bibinfo
  {year} {2009})}\BibitemShut {NoStop}%
\bibitem [{\citenamefont {Tarucha}\ \emph {et~al.}(1995)\citenamefont
  {Tarucha}, \citenamefont {Honda},\ and\ \citenamefont {Saku}}]{Tarucha1995}%
  \BibitemOpen
  \bibfield  {author} {\bibinfo {author} {\bibfnamefont {S.}~\bibnamefont
  {Tarucha}}, \bibinfo {author} {\bibfnamefont {T.}~\bibnamefont {Honda}}, \
  and\ \bibinfo {author} {\bibfnamefont {T.}~\bibnamefont {Saku}},\ }\href
  {\doibase http://dx.doi.org/10.1016/0038-1098(95)00102-6} {\bibfield
  {journal} {\bibinfo  {journal} {Solid State Commun.}\ }\textbf {\bibinfo
  {volume} {94}},\ \bibinfo {pages} {413 } (\bibinfo {year}
  {1995})}\BibitemShut {NoStop}%
\bibitem [{\citenamefont {Yacoby}\ \emph {et~al.}(1996)\citenamefont {Yacoby},
  \citenamefont {Stormer}, \citenamefont {Wingreen}, \citenamefont {Pfeiffer},
  \citenamefont {Baldwin},\ and\ \citenamefont {West}}]{Yacoby1996}%
  \BibitemOpen
  \bibfield  {author} {\bibinfo {author} {\bibfnamefont {A.}~\bibnamefont
  {Yacoby}}, \bibinfo {author} {\bibfnamefont {H.~L.}\ \bibnamefont {Stormer}},
  \bibinfo {author} {\bibfnamefont {N.~S.}\ \bibnamefont {Wingreen}}, \bibinfo
  {author} {\bibfnamefont {L.~N.}\ \bibnamefont {Pfeiffer}}, \bibinfo {author}
  {\bibfnamefont {K.~W.}\ \bibnamefont {Baldwin}}, \ and\ \bibinfo {author}
  {\bibfnamefont {K.~W.}\ \bibnamefont {West}},\ }\href {\doibase
  10.1103/PhysRevLett.77.4612} {\bibfield  {journal} {\bibinfo  {journal}
  {Phys. Rev. Lett.}\ }\textbf {\bibinfo {volume} {77}},\ \bibinfo {pages}
  {4612} (\bibinfo {year} {1996})}\BibitemShut {NoStop}%
\bibitem [{\citenamefont {Thomas}\ \emph {et~al.}(1996)\citenamefont {Thomas},
  \citenamefont {Nicholls}, \citenamefont {Simmons}, \citenamefont {Pepper},
  \citenamefont {Mace},\ and\ \citenamefont {Ritchie}}]{Thomas1996}%
  \BibitemOpen
  \bibfield  {author} {\bibinfo {author} {\bibfnamefont {K.~J.}\ \bibnamefont
  {Thomas}}, \bibinfo {author} {\bibfnamefont {J.~T.}\ \bibnamefont
  {Nicholls}}, \bibinfo {author} {\bibfnamefont {M.~Y.}\ \bibnamefont
  {Simmons}}, \bibinfo {author} {\bibfnamefont {M.}~\bibnamefont {Pepper}},
  \bibinfo {author} {\bibfnamefont {D.~R.}\ \bibnamefont {Mace}}, \ and\
  \bibinfo {author} {\bibfnamefont {D.~A.}\ \bibnamefont {Ritchie}},\ }\href
  {\doibase 10.1103/PhysRevLett.77.135} {\bibfield  {journal} {\bibinfo
  {journal} {Phys. Rev. Lett.}\ }\textbf {\bibinfo {volume} {77}},\ \bibinfo
  {pages} {135} (\bibinfo {year} {1996})}\BibitemShut {NoStop}%
\bibitem [{\citenamefont {Nuttinck}\ \emph {et~al.}(2000)\citenamefont
  {Nuttinck}, \citenamefont {Hashimoto}, \citenamefont {Miyashita},
  \citenamefont {Saku}, \citenamefont {Yamamoto},\ and\ \citenamefont
  {Hirayama}}]{Nuttinck2000}%
  \BibitemOpen
  \bibfield  {author} {\bibinfo {author} {\bibfnamefont {S.}~\bibnamefont
  {Nuttinck}}, \bibinfo {author} {\bibfnamefont {K.}~\bibnamefont {Hashimoto}},
  \bibinfo {author} {\bibfnamefont {S.}~\bibnamefont {Miyashita}}, \bibinfo
  {author} {\bibfnamefont {T.}~\bibnamefont {Saku}}, \bibinfo {author}
  {\bibfnamefont {Y.}~\bibnamefont {Yamamoto}}, \ and\ \bibinfo {author}
  {\bibfnamefont {Y.}~\bibnamefont {Hirayama}},\ }\href {\doibase
  10.1143/JJAP.39.L655} {\bibfield  {journal} {\bibinfo  {journal} {Jpn. J.
  Appl. Phys.}\ }\textbf {\bibinfo {volume} {39}},\ \bibinfo {pages} {L655}
  (\bibinfo {year} {2000})}\BibitemShut {NoStop}%
\bibitem [{\citenamefont {Cronenwett}\ \emph {et~al.}(2002)\citenamefont
  {Cronenwett}, \citenamefont {Lynch}, \citenamefont {Goldhaber-Gordon},
  \citenamefont {Kouwenhoven}, \citenamefont {Marcus}, \citenamefont {Hirose},
  \citenamefont {Wingreen},\ and\ \citenamefont {Umansky}}]{Cronenwett2002}%
  \BibitemOpen
  \bibfield  {author} {\bibinfo {author} {\bibfnamefont {S.~M.}\ \bibnamefont
  {Cronenwett}}, \bibinfo {author} {\bibfnamefont {H.~J.}\ \bibnamefont
  {Lynch}}, \bibinfo {author} {\bibfnamefont {D.}~\bibnamefont
  {Goldhaber-Gordon}}, \bibinfo {author} {\bibfnamefont {L.~P.}\ \bibnamefont
  {Kouwenhoven}}, \bibinfo {author} {\bibfnamefont {C.~M.}\ \bibnamefont
  {Marcus}}, \bibinfo {author} {\bibfnamefont {K.}~\bibnamefont {Hirose}},
  \bibinfo {author} {\bibfnamefont {N.~S.}\ \bibnamefont {Wingreen}}, \ and\
  \bibinfo {author} {\bibfnamefont {V.}~\bibnamefont {Umansky}},\ }\href
  {\doibase 10.1103/PhysRevLett.88.226805} {\bibfield  {journal} {\bibinfo
  {journal} {Phys. Rev. Lett.}\ }\textbf {\bibinfo {volume} {88}},\ \bibinfo
  {pages} {226805} (\bibinfo {year} {2002})}\BibitemShut {NoStop}%
\bibitem [{\citenamefont {Auslaender}\ \emph {et~al.}(2005)\citenamefont
  {Auslaender}, \citenamefont {Steinberg}, \citenamefont {Yacoby},
  \citenamefont {Tserkovnyak}, \citenamefont {Halperin}, \citenamefont
  {Baldwin}, \citenamefont {Pfeiffer},\ and\ \citenamefont
  {West}}]{Auslaender2005}%
  \BibitemOpen
  \bibfield  {author} {\bibinfo {author} {\bibfnamefont {O.~M.}\ \bibnamefont
  {Auslaender}}, \bibinfo {author} {\bibfnamefont {H.}~\bibnamefont
  {Steinberg}}, \bibinfo {author} {\bibfnamefont {A.}~\bibnamefont {Yacoby}},
  \bibinfo {author} {\bibfnamefont {Y.}~\bibnamefont {Tserkovnyak}}, \bibinfo
  {author} {\bibfnamefont {B.~I.}\ \bibnamefont {Halperin}}, \bibinfo {author}
  {\bibfnamefont {K.~W.}\ \bibnamefont {Baldwin}}, \bibinfo {author}
  {\bibfnamefont {L.~N.}\ \bibnamefont {Pfeiffer}}, \ and\ \bibinfo {author}
  {\bibfnamefont {K.~W.}\ \bibnamefont {West}},\ }\href {\doibase
  10.1126/science.1107821} {\bibfield  {journal} {\bibinfo  {journal}
  {Science}\ }\textbf {\bibinfo {volume} {308}},\ \bibinfo {pages} {88}
  (\bibinfo {year} {2005})}\BibitemShut {NoStop}%
\bibitem [{\citenamefont {Jompol}\ \emph {et~al.}(2009)\citenamefont {Jompol},
  \citenamefont {Ford}, \citenamefont {Griffiths}, \citenamefont {Farrer},
  \citenamefont {Jones}, \citenamefont {Anderson}, \citenamefont {Ritchie},
  \citenamefont {Silk},\ and\ \citenamefont {Schofield}}]{Jompol2009}%
  \BibitemOpen
  \bibfield  {author} {\bibinfo {author} {\bibfnamefont {Y.}~\bibnamefont
  {Jompol}}, \bibinfo {author} {\bibfnamefont {C.~J.~B.}\ \bibnamefont {Ford}},
  \bibinfo {author} {\bibfnamefont {J.~P.}\ \bibnamefont {Griffiths}}, \bibinfo
  {author} {\bibfnamefont {I.}~\bibnamefont {Farrer}}, \bibinfo {author}
  {\bibfnamefont {G.~A.~C.}\ \bibnamefont {Jones}}, \bibinfo {author}
  {\bibfnamefont {D.}~\bibnamefont {Anderson}}, \bibinfo {author}
  {\bibfnamefont {D.~A.}\ \bibnamefont {Ritchie}}, \bibinfo {author}
  {\bibfnamefont {T.~W.}\ \bibnamefont {Silk}}, \ and\ \bibinfo {author}
  {\bibfnamefont {A.~J.}\ \bibnamefont {Schofield}},\ }\href {\doibase
  10.1126/science.1171769} {\bibfield  {journal} {\bibinfo  {journal}
  {Science}\ }\textbf {\bibinfo {volume} {325}},\ \bibinfo {pages} {597}
  (\bibinfo {year} {2009})}\BibitemShut {NoStop}%
\bibitem [{\citenamefont {Kane}\ \emph {et~al.}(1998)\citenamefont {Kane},
  \citenamefont {Facer}, \citenamefont {Dzurak}, \citenamefont {Lumpkin},
  \citenamefont {Clark}, \citenamefont {Pfeiffer},\ and\ \citenamefont
  {West}}]{Kane1998}%
  \BibitemOpen
  \bibfield  {author} {\bibinfo {author} {\bibfnamefont {B.~E.}\ \bibnamefont
  {Kane}}, \bibinfo {author} {\bibfnamefont {G.~R.}\ \bibnamefont {Facer}},
  \bibinfo {author} {\bibfnamefont {A.~S.}\ \bibnamefont {Dzurak}}, \bibinfo
  {author} {\bibfnamefont {N.~E.}\ \bibnamefont {Lumpkin}}, \bibinfo {author}
  {\bibfnamefont {R.~G.}\ \bibnamefont {Clark}}, \bibinfo {author}
  {\bibfnamefont {L.~N.}\ \bibnamefont {Pfeiffer}}, \ and\ \bibinfo {author}
  {\bibfnamefont {K.~W.}\ \bibnamefont {West}},\ }\href {\doibase
  10.1063/1.121642} {\bibfield  {journal} {\bibinfo  {journal} {Appl. Phys.
  Lett.}\ }\textbf {\bibinfo {volume} {72}},\ \bibinfo {pages} {3506} (\bibinfo
  {year} {1998})}\BibitemShut {NoStop}%
\bibitem [{\citenamefont {Thomas}\ \emph {et~al.}(2000)\citenamefont {Thomas},
  \citenamefont {Nicholls}, \citenamefont {Pepper}, \citenamefont {Tribe},
  \citenamefont {Simmons},\ and\ \citenamefont {Ritchie}}]{Thomas2000}%
  \BibitemOpen
  \bibfield  {author} {\bibinfo {author} {\bibfnamefont {K.~J.}\ \bibnamefont
  {Thomas}}, \bibinfo {author} {\bibfnamefont {J.~T.}\ \bibnamefont
  {Nicholls}}, \bibinfo {author} {\bibfnamefont {M.}~\bibnamefont {Pepper}},
  \bibinfo {author} {\bibfnamefont {W.~R.}\ \bibnamefont {Tribe}}, \bibinfo
  {author} {\bibfnamefont {M.~Y.}\ \bibnamefont {Simmons}}, \ and\ \bibinfo
  {author} {\bibfnamefont {D.~A.}\ \bibnamefont {Ritchie}},\ }\href {\doibase
  10.1103/PhysRevB.61.R13365} {\bibfield  {journal} {\bibinfo  {journal} {Phys.
  Rev. B}\ }\textbf {\bibinfo {volume} {61}},\ \bibinfo {pages} {R13365}
  (\bibinfo {year} {2000})}\BibitemShut {NoStop}%
\bibitem [{\citenamefont {Reilly}\ \emph {et~al.}(2002)\citenamefont {Reilly},
  \citenamefont {Buehler}, \citenamefont {O'Brien}, \citenamefont {Hamilton},
  \citenamefont {Dzurak}, \citenamefont {Clark}, \citenamefont {Kane},
  \citenamefont {Pfeiffer},\ and\ \citenamefont {West}}]{Reilly2002}%
  \BibitemOpen
  \bibfield  {author} {\bibinfo {author} {\bibfnamefont {D.~J.}\ \bibnamefont
  {Reilly}}, \bibinfo {author} {\bibfnamefont {T.~M.}\ \bibnamefont {Buehler}},
  \bibinfo {author} {\bibfnamefont {J.~L.}\ \bibnamefont {O'Brien}}, \bibinfo
  {author} {\bibfnamefont {A.~R.}\ \bibnamefont {Hamilton}}, \bibinfo {author}
  {\bibfnamefont {A.~S.}\ \bibnamefont {Dzurak}}, \bibinfo {author}
  {\bibfnamefont {R.~G.}\ \bibnamefont {Clark}}, \bibinfo {author}
  {\bibfnamefont {B.~E.}\ \bibnamefont {Kane}}, \bibinfo {author}
  {\bibfnamefont {L.~N.}\ \bibnamefont {Pfeiffer}}, \ and\ \bibinfo {author}
  {\bibfnamefont {K.~W.}\ \bibnamefont {West}},\ }\href {\doibase
  10.1103/PhysRevLett.89.246801} {\bibfield  {journal} {\bibinfo  {journal}
  {Phys. Rev. Lett.}\ }\textbf {\bibinfo {volume} {89}},\ \bibinfo {pages}
  {246801} (\bibinfo {year} {2002})}\BibitemShut {NoStop}%
\bibitem [{\citenamefont {Starikov}\ \emph {et~al.}(2003)\citenamefont
  {Starikov}, \citenamefont {Yakimenko},\ and\ \citenamefont
  {Berggren}}]{Starikov2003}%
  \BibitemOpen
  \bibfield  {author} {\bibinfo {author} {\bibfnamefont {A.~A.}\ \bibnamefont
  {Starikov}}, \bibinfo {author} {\bibfnamefont {I.~I.}\ \bibnamefont
  {Yakimenko}}, \ and\ \bibinfo {author} {\bibfnamefont {K.-F.}\ \bibnamefont
  {Berggren}},\ }\href {\doibase 10.1103/PhysRevB.67.235319} {\bibfield
  {journal} {\bibinfo  {journal} {Phys. Rev. B}\ }\textbf {\bibinfo {volume}
  {67}},\ \bibinfo {pages} {235319} (\bibinfo {year} {2003})}\BibitemShut
  {NoStop}%
\bibitem [{\citenamefont {Havu}\ \emph {et~al.}(2004)\citenamefont {Havu},
  \citenamefont {Puska}, \citenamefont {Nieminen},\ and\ \citenamefont
  {Havu}}]{Havu2004}%
  \BibitemOpen
  \bibfield  {author} {\bibinfo {author} {\bibfnamefont {P.}~\bibnamefont
  {Havu}}, \bibinfo {author} {\bibfnamefont {M.~J.}\ \bibnamefont {Puska}},
  \bibinfo {author} {\bibfnamefont {R.~M.}\ \bibnamefont {Nieminen}}, \ and\
  \bibinfo {author} {\bibfnamefont {V.}~\bibnamefont {Havu}},\ }\href {\doibase
  10.1103/PhysRevB.70.233308} {\bibfield  {journal} {\bibinfo  {journal} {Phys.
  Rev. B}\ }\textbf {\bibinfo {volume} {70}},\ \bibinfo {pages} {233308}
  (\bibinfo {year} {2004})}\BibitemShut {NoStop}%
\bibitem [{\citenamefont {Klironomos}\ \emph {et~al.}(2006)\citenamefont
  {Klironomos}, \citenamefont {Meyer},\ and\ \citenamefont
  {Matveev}}]{Klironomos2006}%
  \BibitemOpen
  \bibfield  {author} {\bibinfo {author} {\bibfnamefont {A.~D.}\ \bibnamefont
  {Klironomos}}, \bibinfo {author} {\bibfnamefont {J.~S.}\ \bibnamefont
  {Meyer}}, \ and\ \bibinfo {author} {\bibfnamefont {K.~A.}\ \bibnamefont
  {Matveev}},\ }\href {http://stacks.iop.org/0295-5075/74/i=4/a=679} {\bibfield
   {journal} {\bibinfo  {journal} {Europhys. Lett.}\ }\textbf {\bibinfo
  {volume} {74}},\ \bibinfo {pages} {679} (\bibinfo {year} {2006})}\BibitemShut
  {NoStop}%
\bibitem [{\citenamefont {Rokhinson}\ \emph {et~al.}(2006)\citenamefont
  {Rokhinson}, \citenamefont {Pfeiffer},\ and\ \citenamefont
  {West}}]{Rokhinson2006}%
  \BibitemOpen
  \bibfield  {author} {\bibinfo {author} {\bibfnamefont {L.~P.}\ \bibnamefont
  {Rokhinson}}, \bibinfo {author} {\bibfnamefont {L.~N.}\ \bibnamefont
  {Pfeiffer}}, \ and\ \bibinfo {author} {\bibfnamefont {K.~W.}\ \bibnamefont
  {West}},\ }\href {\doibase 10.1103/PhysRevLett.96.156602} {\bibfield
  {journal} {\bibinfo  {journal} {Phys. Rev. Lett.}\ }\textbf {\bibinfo
  {volume} {96}},\ \bibinfo {pages} {156602} (\bibinfo {year}
  {2006})}\BibitemShut {NoStop}%
\bibitem [{\citenamefont {Berggren}\ and\ \citenamefont
  {Yakimenko}(2008)}]{Berggren2008}%
  \BibitemOpen
  \bibfield  {author} {\bibinfo {author} {\bibfnamefont {K.-F.}\ \bibnamefont
  {Berggren}}\ and\ \bibinfo {author} {\bibfnamefont {I.~I.}\ \bibnamefont
  {Yakimenko}},\ }\href {\doibase 10.1088/0953-8984/20/16/164203} {\bibfield
  {journal} {\bibinfo  {journal} {J. Phys.: Condens. Matter}\ }\textbf
  {\bibinfo {volume} {20}},\ \bibinfo {pages} {164203} (\bibinfo {year}
  {2008})}\BibitemShut {NoStop}%
\bibitem [{\citenamefont {Micolich}(2011)}]{Micolich2011}%
  \BibitemOpen
  \bibfield  {author} {\bibinfo {author} {\bibfnamefont {A.~P.}\ \bibnamefont
  {Micolich}},\ }\href {\doibase 10.1088/0953-8984/23/44/443201} {\bibfield
  {journal} {\bibinfo  {journal} {J. Phys.: Condens. Matter}\ }\textbf
  {\bibinfo {volume} {23}},\ \bibinfo {pages} {443201} (\bibinfo {year}
  {2011})}\BibitemShut {NoStop}%
\bibitem [{\citenamefont {Lind}\ \emph {et~al.}(2011)\citenamefont {Lind},
  \citenamefont {Yakimenko},\ and\ \citenamefont {Berggren}}]{Lind2011}%
  \BibitemOpen
  \bibfield  {author} {\bibinfo {author} {\bibfnamefont {H.}~\bibnamefont
  {Lind}}, \bibinfo {author} {\bibfnamefont {I.~I.}\ \bibnamefont {Yakimenko}},
  \ and\ \bibinfo {author} {\bibfnamefont {K.-F.}\ \bibnamefont {Berggren}},\
  }\href {\doibase 10.1103/PhysRevB.83.075308} {\bibfield  {journal} {\bibinfo
  {journal} {Phys. Rev. B}\ }\textbf {\bibinfo {volume} {83}},\ \bibinfo
  {pages} {075308} (\bibinfo {year} {2011})}\BibitemShut {NoStop}%
\bibitem [{\citenamefont {Hirose}\ \emph {et~al.}(2003)\citenamefont {Hirose},
  \citenamefont {Meir},\ and\ \citenamefont {Wingreen}}]{Hirose2003}%
  \BibitemOpen
  \bibfield  {author} {\bibinfo {author} {\bibfnamefont {K.}~\bibnamefont
  {Hirose}}, \bibinfo {author} {\bibfnamefont {Y.}~\bibnamefont {Meir}}, \ and\
  \bibinfo {author} {\bibfnamefont {N.~S.}\ \bibnamefont {Wingreen}},\ }\href
  {\doibase 10.1103/PhysRevLett.90.026804} {\bibfield  {journal} {\bibinfo
  {journal} {Phys. Rev. Lett.}\ }\textbf {\bibinfo {volume} {90}},\ \bibinfo
  {pages} {026804} (\bibinfo {year} {2003})}\BibitemShut {NoStop}%
\bibitem [{\citenamefont {Rejec}\ and\ \citenamefont {Meir}(2006)}]{Rejec2006}%
  \BibitemOpen
  \bibfield  {author} {\bibinfo {author} {\bibfnamefont {T.}~\bibnamefont
  {Rejec}}\ and\ \bibinfo {author} {\bibfnamefont {Y.}~\bibnamefont {Meir}},\
  }\href {\doibase 10.1038/nature05054} {\bibfield  {journal} {\bibinfo
  {journal} {Nature}\ }\textbf {\bibinfo {volume} {442}},\ \bibinfo {pages}
  {900} (\bibinfo {year} {2006})}\BibitemShut {NoStop}%
\bibitem [{\citenamefont {Lieb}\ and\ \citenamefont
  {Mattis}(1962)}]{Lieb1962a}%
  \BibitemOpen
  \bibfield  {author} {\bibinfo {author} {\bibfnamefont {E.}~\bibnamefont
  {Lieb}}\ and\ \bibinfo {author} {\bibfnamefont {D.}~\bibnamefont {Mattis}},\
  }\href {\doibase 10.1103/PhysRev.125.164} {\bibfield  {journal} {\bibinfo
  {journal} {Phys. Rev.}\ }\textbf {\bibinfo {volume} {125}},\ \bibinfo {pages}
  {164} (\bibinfo {year} {1962})}\BibitemShut {NoStop}%
\bibitem [{\citenamefont {Debray}\ \emph {et~al.}(2009)\citenamefont {Debray},
  \citenamefont {Rahman}, \citenamefont {Wan}, \citenamefont {Newrock},
  \citenamefont {Cahay}, \citenamefont {Ngo}, \citenamefont {Ulloa},
  \citenamefont {Herbert}, \citenamefont {Muhammad},\ and\ \citenamefont
  {Johnson}}]{Debray2009}%
  \BibitemOpen
  \bibfield  {author} {\bibinfo {author} {\bibfnamefont {P.}~\bibnamefont
  {Debray}}, \bibinfo {author} {\bibfnamefont {S.~M.~S.}\ \bibnamefont
  {Rahman}}, \bibinfo {author} {\bibfnamefont {J.}~\bibnamefont {Wan}},
  \bibinfo {author} {\bibfnamefont {R.~S.}\ \bibnamefont {Newrock}}, \bibinfo
  {author} {\bibfnamefont {M.}~\bibnamefont {Cahay}}, \bibinfo {author}
  {\bibfnamefont {A.~T.}\ \bibnamefont {Ngo}}, \bibinfo {author} {\bibfnamefont
  {S.~E.}\ \bibnamefont {Ulloa}}, \bibinfo {author} {\bibfnamefont {S.~T.}\
  \bibnamefont {Herbert}}, \bibinfo {author} {\bibfnamefont {M.}~\bibnamefont
  {Muhammad}}, \ and\ \bibinfo {author} {\bibfnamefont {M.}~\bibnamefont
  {Johnson}},\ }\href
  {http://www.nature.com/nnano/journal/v4/n11/abs/nnano.2009.240.html}
  {\bibfield  {journal} {\bibinfo  {journal} {Nat. Nanotechnol.}\ }\textbf
  {\bibinfo {volume} {4}},\ \bibinfo {pages} {759} (\bibinfo {year}
  {2009})}\BibitemShut {NoStop}%
\bibitem [{\citenamefont {Wan}\ \emph {et~al.}(2011)\citenamefont {Wan},
  \citenamefont {Cahay}, \citenamefont {Debray},\ and\ \citenamefont
  {Newrock}}]{Wan2011}%
  \BibitemOpen
  \bibfield  {author} {\bibinfo {author} {\bibfnamefont {J.}~\bibnamefont
  {Wan}}, \bibinfo {author} {\bibfnamefont {M.}~\bibnamefont {Cahay}}, \bibinfo
  {author} {\bibfnamefont {P.}~\bibnamefont {Debray}}, \ and\ \bibinfo {author}
  {\bibfnamefont {R.~S.}\ \bibnamefont {Newrock}},\ }\href {\doibase
  http://dx.doi.org/10.1166/jno.2011.1139} {\bibfield  {journal} {\bibinfo
  {journal} {J. Nanoelectron. Optoelectron.}\ }\textbf {\bibinfo {volume}
  {6}},\ \bibinfo {pages} {95} (\bibinfo {year} {2011})}\BibitemShut {NoStop}%
\bibitem [{\citenamefont {Sánchez}\ and\ \citenamefont
  {Leburton}(2013)}]{Sanchez2013}%
  \BibitemOpen
  \bibfield  {author} {\bibinfo {author} {\bibfnamefont {A.~X.}\ \bibnamefont
  {Sánchez}}\ and\ \bibinfo {author} {\bibfnamefont {J.-P.}\ \bibnamefont
  {Leburton}},\ }\href {\doibase 10.1103/PhysRevB.88.075305} {\bibfield
  {journal} {\bibinfo  {journal} {Phys. Rev. B}\ }\textbf {\bibinfo {volume}
  {88}},\ \bibinfo {pages} {075305} (\bibinfo {year} {2013})}\BibitemShut
  {NoStop}%
\bibitem [{\citenamefont {Hess}(2000)}]{Hess2000}%
  \BibitemOpen
  \bibfield  {author} {\bibinfo {author} {\bibfnamefont {K.}~\bibnamefont
  {Hess}},\ }\href
  {http://ieeexplore.ieee.org/xpl/bkabstractplus.jsp?bkn=5265897} {\emph
  {\bibinfo {title} {Advanced Theory of Semiconductor Devices}}}\ (\bibinfo
  {publisher} {Wiley-IEEE Press},\ \bibinfo {address} {New York},\ \bibinfo
  {year} {2000})\BibitemShut {NoStop}%
\bibitem [{\citenamefont {Graham}\ \emph {et~al.}(2007)\citenamefont {Graham},
  \citenamefont {Sawkey}, \citenamefont {Pepper}, \citenamefont {Simmons},\
  and\ \citenamefont {Ritchie}}]{Graham2007}%
  \BibitemOpen
  \bibfield  {author} {\bibinfo {author} {\bibfnamefont {A.~C.}\ \bibnamefont
  {Graham}}, \bibinfo {author} {\bibfnamefont {D.~L.}\ \bibnamefont {Sawkey}},
  \bibinfo {author} {\bibfnamefont {M.}~\bibnamefont {Pepper}}, \bibinfo
  {author} {\bibfnamefont {M.~Y.}\ \bibnamefont {Simmons}}, \ and\ \bibinfo
  {author} {\bibfnamefont {D.~A.}\ \bibnamefont {Ritchie}},\ }\href {\doibase
  10.1103/PhysRevB.75.035331} {\bibfield  {journal} {\bibinfo  {journal} {Phys.
  Rev. B}\ }\textbf {\bibinfo {volume} {75}},\ \bibinfo {pages} {035331}
  (\bibinfo {year} {2007})}\BibitemShut {NoStop}%
\bibitem [{\citenamefont {Berggren}\ \emph {et~al.}(2005)\citenamefont
  {Berggren}, \citenamefont {Jaksch},\ and\ \citenamefont
  {Yakimenko}}]{Berggren2005}%
  \BibitemOpen
  \bibfield  {author} {\bibinfo {author} {\bibfnamefont {K.-F.}\ \bibnamefont
  {Berggren}}, \bibinfo {author} {\bibfnamefont {P.}~\bibnamefont {Jaksch}}, \
  and\ \bibinfo {author} {\bibfnamefont {I.}~\bibnamefont {Yakimenko}},\ }\href
  {\doibase 10.1103/PhysRevB.71.115303} {\bibfield  {journal} {\bibinfo
  {journal} {Phys. Rev. B}\ }\textbf {\bibinfo {volume} {71}},\ \bibinfo
  {pages} {115303} (\bibinfo {year} {2005})}\BibitemShut {NoStop}%
\bibitem [{\citenamefont {Lassl}\ \emph {et~al.}(2007)\citenamefont {Lassl},
  \citenamefont {Schlagheck},\ and\ \citenamefont {Richter}}]{Lassl2007}%
  \BibitemOpen
  \bibfield  {author} {\bibinfo {author} {\bibfnamefont {A.}~\bibnamefont
  {Lassl}}, \bibinfo {author} {\bibfnamefont {P.}~\bibnamefont {Schlagheck}}, \
  and\ \bibinfo {author} {\bibfnamefont {K.}~\bibnamefont {Richter}},\ }\href
  {\doibase 10.1103/PhysRevB.75.045346} {\bibfield  {journal} {\bibinfo
  {journal} {Phys. Rev. B}\ }\textbf {\bibinfo {volume} {75}},\ \bibinfo
  {pages} {045346} (\bibinfo {year} {2007})}\BibitemShut {NoStop}%
\bibitem [{\citenamefont {Kristensen}\ \emph {et~al.}(2000)\citenamefont
  {Kristensen}, \citenamefont {Bruus}, \citenamefont {Hansen}, \citenamefont
  {Jensen}, \citenamefont {Lindelof}, \citenamefont {Marckmann}, \citenamefont
  {Nyg\aa{}rd}, \citenamefont {S\o{}rensen}, \citenamefont {Beuscher},
  \citenamefont {Forchel},\ and\ \citenamefont {Michel}}]{Kristensen2000}%
  \BibitemOpen
  \bibfield  {author} {\bibinfo {author} {\bibfnamefont {A.}~\bibnamefont
  {Kristensen}}, \bibinfo {author} {\bibfnamefont {H.}~\bibnamefont {Bruus}},
  \bibinfo {author} {\bibfnamefont {A.~E.}\ \bibnamefont {Hansen}}, \bibinfo
  {author} {\bibfnamefont {J.~B.}\ \bibnamefont {Jensen}}, \bibinfo {author}
  {\bibfnamefont {P.~E.}\ \bibnamefont {Lindelof}}, \bibinfo {author}
  {\bibfnamefont {C.~J.}\ \bibnamefont {Marckmann}}, \bibinfo {author}
  {\bibfnamefont {J.}~\bibnamefont {Nyg\aa{}rd}}, \bibinfo {author}
  {\bibfnamefont {C.~B.}\ \bibnamefont {S\o{}rensen}}, \bibinfo {author}
  {\bibfnamefont {F.}~\bibnamefont {Beuscher}}, \bibinfo {author}
  {\bibfnamefont {A.}~\bibnamefont {Forchel}}, \ and\ \bibinfo {author}
  {\bibfnamefont {M.}~\bibnamefont {Michel}},\ }\href {\doibase
  10.1103/PhysRevB.62.10950} {\bibfield  {journal} {\bibinfo  {journal} {Phys.
  Rev. B}\ }\textbf {\bibinfo {volume} {62}},\ \bibinfo {pages} {10950}
  (\bibinfo {year} {2000})}\BibitemShut {NoStop}%
\bibitem [{\citenamefont {Imambekov}\ \emph {et~al.}(2012)\citenamefont
  {Imambekov}, \citenamefont {Schmidt},\ and\ \citenamefont
  {Glazman}}]{Imambekov2012}%
  \BibitemOpen
  \bibfield  {author} {\bibinfo {author} {\bibfnamefont {A.}~\bibnamefont
  {Imambekov}}, \bibinfo {author} {\bibfnamefont {T.~L.}\ \bibnamefont
  {Schmidt}}, \ and\ \bibinfo {author} {\bibfnamefont {L.~I.}\ \bibnamefont
  {Glazman}},\ }\href {\doibase 10.1103/RevModPhys.84.1253} {\bibfield
  {journal} {\bibinfo  {journal} {Rev. Mod. Phys.}\ }\textbf {\bibinfo {volume}
  {84}},\ \bibinfo {pages} {1253} (\bibinfo {year} {2012})}\BibitemShut
  {NoStop}%
\bibitem [{\citenamefont {Barak}\ \emph {et~al.}(2010)\citenamefont {Barak},
  \citenamefont {Steinberg}, \citenamefont {Pfeiffer}, \citenamefont {West},
  \citenamefont {Glazman}, \citenamefont {von Oppen},\ and\ \citenamefont
  {Yacoby}}]{Barak2010}%
  \BibitemOpen
  \bibfield  {author} {\bibinfo {author} {\bibfnamefont {G.}~\bibnamefont
  {Barak}}, \bibinfo {author} {\bibfnamefont {H.}~\bibnamefont {Steinberg}},
  \bibinfo {author} {\bibfnamefont {L.~N.}\ \bibnamefont {Pfeiffer}}, \bibinfo
  {author} {\bibfnamefont {K.~W.}\ \bibnamefont {West}}, \bibinfo {author}
  {\bibfnamefont {L.}~\bibnamefont {Glazman}}, \bibinfo {author} {\bibfnamefont
  {F.}~\bibnamefont {von Oppen}}, \ and\ \bibinfo {author} {\bibfnamefont
  {A.}~\bibnamefont {Yacoby}},\ }\href {\doibase 10.1038/nphys1678} {\bibfield
  {journal} {\bibinfo  {journal} {Nat. Phys.}\ }\textbf {\bibinfo {volume}
  {6}},\ \bibinfo {pages} {489} (\bibinfo {year} {2010})}\BibitemShut {NoStop}%
\bibitem [{\citenamefont {Deshpande}\ \emph {et~al.}(2010)\citenamefont
  {Deshpande}, \citenamefont {Bockrath}, \citenamefont {Glazman},\ and\
  \citenamefont {Yacoby}}]{Deshpande2010}%
  \BibitemOpen
  \bibfield  {author} {\bibinfo {author} {\bibfnamefont {V.~V.}\ \bibnamefont
  {Deshpande}}, \bibinfo {author} {\bibfnamefont {M.}~\bibnamefont {Bockrath}},
  \bibinfo {author} {\bibfnamefont {L.~I.}\ \bibnamefont {Glazman}}, \ and\
  \bibinfo {author} {\bibfnamefont {A.}~\bibnamefont {Yacoby}},\ }\href
  {http://dx.doi.org/10.1038/nature08918} {\bibfield  {journal} {\bibinfo
  {journal} {Nature}\ }\textbf {\bibinfo {volume} {464}},\ \bibinfo {pages}
  {209} (\bibinfo {year} {2010})}\BibitemShut {NoStop}%
\end{thebibliography}%

\pagebreak{}

\setcounter{equation}{0}
\setcounter{figure}{0}
\renewcommand{\theequation}{S\arabic{equation}}
\renewcommand{\thefigure}{S\arabic{figure}}

\section*{Supplementary Methods}

\subsection*{Variational calculation of the 1D energy-momentum relation}

\begin{comment}
Add \textquotedbl{}\textbackslash{}pagestyle\{empty\}\textquotedbl{}
(without quotation marks) to the ERT box above to remove numbering
in this section.
\end{comment}
In a longitudinal magnetic field $\vec{B}=B_{x}\hat{x}$, with the
gauge set to $\vec{A}=B_{x}y\hat{z}$, the Schrödinger equation within
the unrestricted Hartree-Fock approximation\citep{Sanchez2013} reads
\begin{multline}
-\frac{\hbar^{2}}{2m^{*}}\nabla^{2}\psi_{\left\{ i\right\} }\left(\vec{r}\right)+\left[U_{\mathrm{conf}}\left(y,z\right)+\frac{1}{2}m^{*}\omega_{B}^{2}y^{2}+i\hbar\omega_{B}y\frac{\partial}{\partial z}+U_{\mathrm{el}}\left(\vec{r}\right)\right]\psi_{\left\{ i\right\} }\left(\vec{r}\right)\\
+\hat{U}_{\mathrm{exch}}\left[\psi_{\left\{ i\right\} }\left(\vec{r}\right)\right]+U_{Z}\psi_{\left\{ i\right\} }\left(\vec{r}\right)=E_{\left\{ i\right\} }\psi_{\left\{ i\right\} }\left(\vec{r}\right)\label{eq:SM__Schr}
\end{multline}
Here, $\left\{ i\right\} =\left\{ i_{x},i_{y},i_{z},\sigma_{i}=\pm\frac{1}{2}\right\} $
are the quantum numbers associated with the eigenenergies $E_{\left\{ i\right\} }$,
$\omega_{B}=\frac{qB_{0}}{m^{*}}$ is the cyclotron frequency, $q$
is the electron charge, $m^{*}$ is the effective mass, $U_{\mathrm{conf}}\left(y,z\right)=\frac{1}{2}m^{*}\omega_{y}^{2}y^{2}+\frac{1}{2}m^{*}\omega_{z}^{2}z^{2}$
is the lateral confinement potential, $U_{Z}=g\mu_{B}B_{x}\sigma$
is the Zeeman energy term, $g$ is the effective electron $g$-factor,
$\mu_{B}=\frac{q\hbar}{2m^{*}}$ is the Bohr magneton in the wire,
and $U_{\mathrm{el}}$ and $\hat{U}_{\mathrm{exch}}$ are the Hartree
and exchange terms, respectively\citep{Sanchez2013}.

In the extreme quantum limit, i.e. when only the lowest-energy subband
of the confinement potential is occupied, we take the expectation
value of the left-hand side of Eq. (\ref{eq:SM__Schr}), using the
trial wavefunction 
\begin{equation}
\psi_{k_{x}}\left(\vec{r}\right)=\frac{1}{\sqrt{L_{x}}}\mathrm{e}^{ik_{x}x}\left(\frac{a^{1/2}}{\pi^{1/4}}e^{-a^{2}y^{2}/2}\right)\left(\frac{b^{1/2}}{\pi^{1/4}}e^{-b^{2}z^{2}/2}\right)\label{eq:SM__wavefunction}
\end{equation}
where $a=\sqrt{\frac{m^{*}\omega_{y}}{\hbar}}$ and $b=\sqrt{\frac{m^{*}\omega_{z}}{\hbar}}$.
In the absence of electron-electron interactions and for zero magnetic
field, this wavefunction corresponds to the exact ground state of
Eq. (\ref{eq:SM__Schr}). Then, one obtains the following expression
for the single-particle energy: 
\begin{equation}
E\left(k_{x},\sigma\right)=\frac{\hbar^{2}k_{x}^{2}}{2m^{*}}+\frac{1}{2}\hbar\omega_{y}+\frac{1}{2}\hbar\omega_{z}+\frac{1}{4}\left(\frac{\omega_{B}}{\omega_{y}}\right)\hbar\omega_{B}+g\mu_{B}B_{x}\sigma+U_{\mathrm{el}}\left[n_{0}\right]+U_{\mathrm{exch}}\left(k_{x},\sigma\right)\label{eq:SM__E__kx_sigma__methods}
\end{equation}
The expectation values $U_{\mathrm{el}}$ and $U_{\mathrm{exch}}$
are\citep{Sanchez2013} 
\[
U_{\mathrm{el}}\left[n_{0}\right]=\frac{q^{2}\zeta_{ab}\left(0\right)}{16\pi\epsilon}n_{0}
\]
\[
U_{\mathrm{exch}}\left(k_{x},\sigma\right)=-\frac{q^{2}}{32\pi^{2}\epsilon}\int_{-\infty}^{+\infty}dp_{x}\,\zeta_{ab}\left(p_{x}-k_{x}\right)f_{T}\left[E\left(p_{x},\sigma\right)\right]
\]

where $n_{0}=\frac{1}{2\pi}\sum\limits _{\sigma}\int_{-\infty}^{+\infty}dp\, f_{T}\left[E\left(p,\sigma\right)\right]=\sum\limits _{\sigma}n_{\sigma}$
is the total electron density in the wire; $f_{T}\left[E\right]=\left\{ 1+\exp\left[\left(E-\mu\right)/k_{B}T\right]\right\} ^{-1}$
is the Fermi-Dirac distribution for a chemical potential $\mu$; and
$\zeta_{ab}\left(p\right)$, a form function which is specific to
the wavefunction shown in Eq. (\ref{eq:SM__wavefunction}), is 
\begin{eqnarray}
\zeta_{ab}\left(p\right) & = & 8\left(\frac{a}{b}\right)\int_{0}^{+\infty}dt\,\frac{t\exp\left(-t^{2}/2\right)}{\sqrt{\left(\frac{p}{b}\right)^{2}+t^{2}}\sqrt{\left(\frac{p}{b}\right)^{2}+\left(\frac{a}{b}\right)^{2}t^{2}}}\qquad;\label{eq:SM__zeta_ab}
\end{eqnarray}

$\zeta_{ab}\left(p\right)$ is a monotonically-decreasing function
of $p$ and it increases with increasing $a$ and $b$ (i.e. with
stronger confinement strengths). A plot of $\zeta_{ab}\left(p\right)$
for different values of $a$ and $b$ is shown in Supplementary Figure
\ref{fig:S1}.

At zero temperature, $\mu$ is equal to the Fermi energy $E_{f}$
and $f_{T}\left(E\right)=\theta\left(E_{f}-E\left(k\right)\right)=\theta\left(k_{f\left(\sigma\right)}-\left|k\right|\right)$,
where $k_{f\left(\sigma\right)}$ is the spin-dependent Fermi wavevector
such that $E\left(k_{f\left(\sigma\right)}\right)=E_{f}$. Then, the
exchange term turns into
\[
U_{\mathrm{exch}}\left(k_{x},\sigma\right)=-\frac{q^{2}}{32\pi^{2}\epsilon}\int_{-k_{f\left(\sigma\right)}+k_{x}}^{+k_{f\left(\sigma\right)}+k_{x}}dp\,\zeta_{ab}\left(p\right)
\]

\subsection*{Calculation of $\Delta n_{\sigma}$, $n_{0}^{\,\mathrm{onset}}$
and $n_{0}^{\,\mathrm{full}}$}

At $T=0$, when the concentrations of both spin-up and spin-down electrons
are positive, they are directly proportional to a spin-dependent Fermi
wavevector via the relation $k_{f\left(\sigma\right)}=\pi n_{\sigma}$.
Thus, the Fermi energy $E_{f}=E\left(k_{f\left(\sigma\right)}\right)$
can be written in terms of $n_{\sigma}$ and $n_{0}$:
\begin{eqnarray*}
E_{f}\left(n_{\sigma}\right) & = & \frac{\hbar^{2}\pi^{2}n_{\sigma}^{2}}{2m^{*}}+\frac{1}{2}\hbar\omega_{y}+\frac{1}{2}\hbar\omega_{z}+\frac{1}{4}\left(\frac{\omega_{B}}{\omega_{y}}\right)\hbar\omega_{B}\\
 &  & +g\mu_{B}B_{x}\sigma+\frac{q^{2}\zeta_{ab}\left(0\right)}{16\pi\epsilon}n_{0}-\frac{q^{2}}{32\pi^{2}\epsilon}\int_{-\pi n_{0}/2+\pi n_{\sigma}}^{+\pi n_{0}/2+\pi n_{\sigma}}dp\,\zeta_{ab}\left(p\right)
\end{eqnarray*}

$n_{\sigma}$ can be written in terms of the average concentration
$n_{0}/2$ and the deviation from this average value $\Delta n_{\sigma}\equiv n_{\sigma}-n_{0}/2$
(so $\Delta n_{-\sigma}=-\Delta n_{\sigma}$). Then, since $E_{f}$
is spin-independent, we get $E_{f}\left(n_{\sigma}\right)=E_{f}\left(n_{-\sigma}\right)$.
From this equation one obtains the following expression, which corresponds
to Equation (\ref{eq:n0_Dns}) in the main text:
\begin{equation}
\frac{\pi^{2}}{2}n_{0}\Delta n_{\sigma}+\sigma\frac{qgB_{x}}{2\hbar}-\frac{q^{2}m^{*}}{64\pi^{2}\epsilon\hbar^{2}}\int_{2\pi\left(n_{0}/2-\Delta n_{\sigma}\right)}^{2\pi\left(n_{0}/2+\Delta n_{\sigma}\right)}dp\,\zeta_{ab}\left(p\right)\equiv F\left(\Delta n_{\sigma};n_{0}\right)=0\label{eq:SM__F_Dns_n0}
\end{equation}

For non-interacting electrons (i.e. when the overlap function $\zeta_{ab}\left(p\right)$
vanishes), the magnetic field induces a Zeeman splitting between spin-up
and spin-down electrons. For $n_{0}\geq n_{B}\equiv\sqrt{\frac{q\left|gB_{x}\right|}{\pi^{2}\hbar}}$,
the (unique) solution to Eq. (\ref{eq:SM__F_Dns_n0}) reduces to $n_{\sigma}=\frac{n_{0}}{2}\left[1-\left(\frac{n_{B}}{n_{0}}\right)^{2}\mathrm{sign}\left(gB_{x}\sigma\right)\right]$.
However, for $n_{0}<n_{B}$, if $\mathrm{sign}\left(gB_{x}\right)>\left(<\right)0$,
this last equation yields $n_{\uparrow\left(\downarrow\right)}<0$,
which must then be set to zero, so the wire exhibits full spin-down
(up) polarization ($n_{\downarrow\left(\uparrow\right)}=n_{0}$).

In the general case of interacting electrons ($\zeta_{ab}\left(p\right)\neq0$),
and if the concentration is low, Eq. (\ref{eq:SM__F_Dns_n0}) yields
only one solution for $\Delta n_{\sigma}$ in terms of $n_{0}$. However,
if $n_{0}$ exceeds a certain threshold value $n_{0}^{\,\mathrm{onset}}$
then two additional solutions, or spin configurations, are possible.
This is illustrated in Supplementary Figure \ref{fig:S2}, which is
a plot of $F\left(\Delta n_{\sigma};n_{0}\right)$ (the left-hand
side of Eq. (\ref{eq:SM__F_Dns_n0})) versus $\Delta n_{\uparrow}$
for a few values of $n_{0}$. As in Figure \ref{fig:P__vs__n0} in
the main text, $B_{x}=\unit[1]{T}$ and $\hbar\omega_{y}=\hbar\omega_{z}=\unit[2]{meV}$,
yielding $n_{0}^{\,\mathrm{onset}}=\unit[1.94\times10^{5}]{cm^{-1}}$.
The solutions to Eq. (\ref{eq:SM__F_Dns_n0}) correspond to the crossings
of the curves with the $x$-axis (dotted line). When the concentration
is below $n_{0}^{\,\mathrm{onset}}$ (dashed line), where is a single
positive solution for $\Delta n_{\uparrow}$, which corresponds to
the ``$\uparrow$'' configuration in Fig. \ref{fig:P__vs__n0}.
Meanwhile, when $n_{0}\geq n_{0}^{\,\mathrm{onset}}$ (dash-dotted
line), two additional negative solutions are possible; they correspond
to the ``$\downarrow$'' and ``$\downarrow^{*}$'' configurations
of Fig. \ref{fig:P__vs__n0}. For $n_{0}=n_{0}^{\,\mathrm{onset}}$
(solid line), the slope of the function $F\left(\Delta n_{\uparrow};n_{0}\right)$
is zero at the value of $\Delta n_{\uparrow}$ for which the negative
solution emerges. Thus, $n_{0}^{\,\mathrm{onset}}$ and the corresponding
$\Delta n_{\sigma}$ are the solutions of the pair of equations $F\left(\Delta n_{\sigma};n_{0}\right)=0$
and $\frac{\partial F\left(\Delta n_{\sigma};n_{0}\right)}{\partial\Delta n_{\sigma}}=0$.

In the limit $\left|\Delta n_{\sigma}\right|\ll n_{0}$, one can expand
the integral in Eq. (\ref{eq:SM__F_Dns_n0}) as a series in $\Delta n_{\sigma}$.
Up to the third order, this results in a cubic equation:
\[
\frac{\pi^{2}}{2}n_{0}\Delta n_{\sigma}+\sigma\frac{qgB_{x}}{2\hbar}-\frac{q^{2}m^{*}}{16\pi^{2}\epsilon\hbar^{2}}\left\{ \pi\Delta n_{\sigma}\zeta_{ab}\left(\pi n_{0}\right)+\frac{2}{3}\left(\pi\Delta n_{\sigma}\right)^{3}\zeta_{ab}^{\prime\prime}\left(\pi n_{0}\right)+\ldots\right\} =0
\]

Here, $\zeta_{ab}^{\prime\prime}\left(\pi n_{0}\right)=\left[\frac{d^{2}\zeta_{ab}\left(p\right)}{dp^{2}}\right]_{p=\pi n_{0}}$.
This leads to the approximate expression for the non-zero values of
$\Delta n_{\sigma}$ when $B_{x}=0$ (Eq. (\ref{eq:Dns__v__n0__B0__approx})
in the main text):
\[
\Delta n_{\sigma}\approx\pm\sqrt{\frac{3}{\zeta_{ab}^{\prime\prime}\left(\pi n_{0}\right)}\left[a^{*}n_{0}-\frac{\zeta_{ab}\left(\pi n_{0}\right)}{2\pi^{2}}\right]}
\]
where $a^{*}\equiv\frac{4\pi\epsilon\hbar^{2}}{q^{2}m^{*}}$ is the
effective Bohr radius. However, this solution is only valid if the
argument inside the square root is non-negative, which means that
the minimum concentration $n_{0\left(B_{x}=0\right)}^{\,\mathrm{onset}}$
for the emergence of the non-zero solution must satisfy the identity
\[
\frac{\zeta_{ab}\left(\pi n_{0\left(B_{x}=0\right)}^{\,\mathrm{onset}}\right)}{\pi n_{0\left(B_{x}=0\right)}^{\,\mathrm{onset}}}=2\pi a^{*}
\]

As $n_{0}$ increases above $n_{0}^{\,\mathrm{onset}}$, $\Delta n_{\sigma}$
will continue to increase until the wire is completely spin-polarized
at a concentration $n_{0}^{\,\mathrm{full}}$. Setting $\Delta n_{\sigma}=\pm n_{0}/2$
in Eq. (\ref{eq:SM__F_Dns_n0}) gives the following equation for $n_{0}^{\,\mathrm{full}}$,
the value of $n_{0}$ for which full spin polarization is achieved:
\[
\frac{\pi^{2}}{4}\left(n_{0}^{\,\mathrm{full}}\right)^{2}-\frac{q^{2}m^{*}}{64\pi^{2}\epsilon\hbar^{2}}\int_{0}^{2\pi n_{0}^{\,\mathrm{full}}}dp\,\zeta_{ab}\left(p\right)=-\mathrm{sign}\left(\sigma\Delta n_{\sigma}\right)\frac{qgB_{x}}{4\hbar}
\]

\begin{figure}[h]
\noindent \begin{centering}
\includegraphics[width=3in]{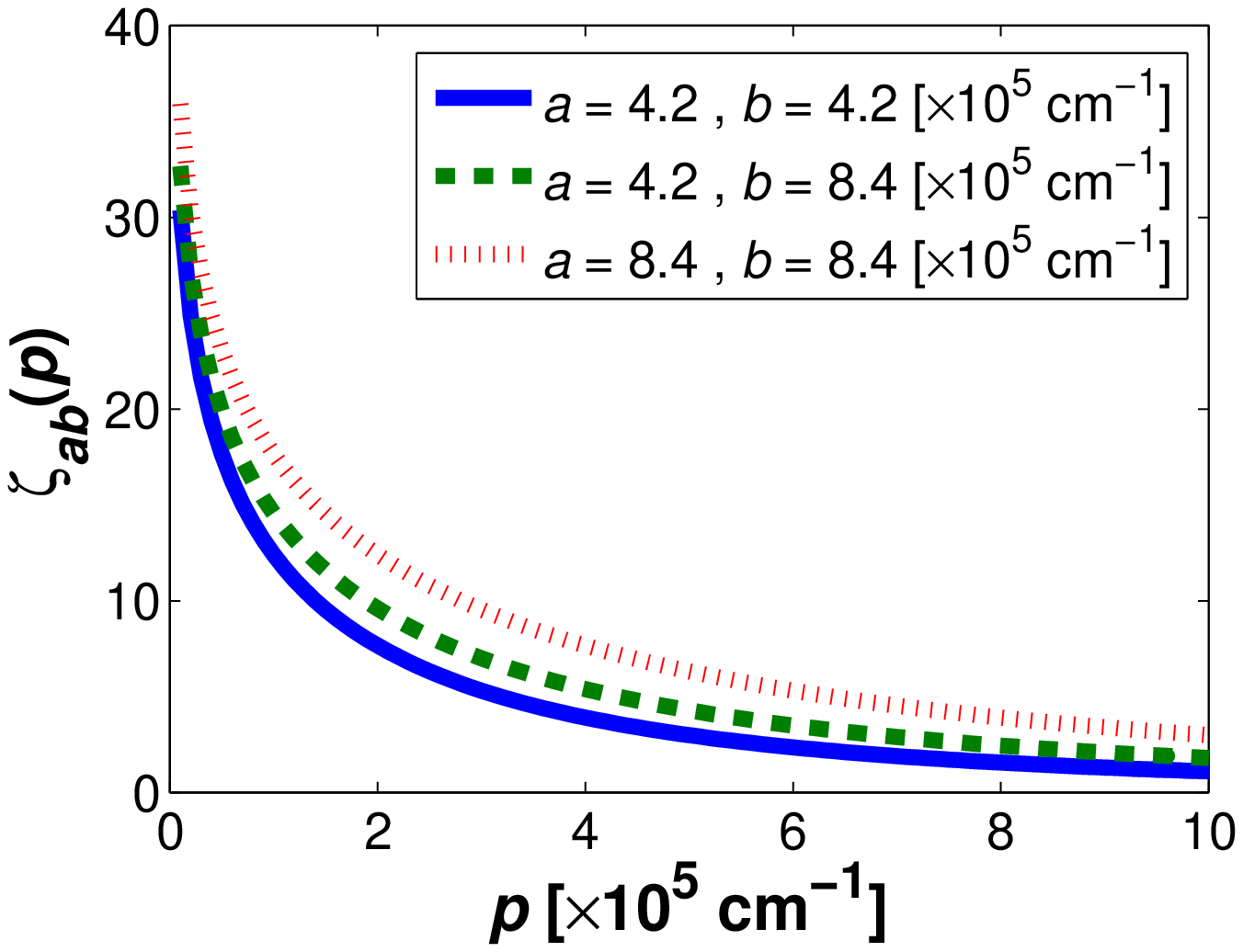}
\par\end{centering}

\protect\caption{\label{fig:S1}$\zeta_{ab}\left(p\right)$ as a function of $p$ for
different values of $a$ and $b$.}
\end{figure}

\begin{figure}[h]
\noindent \begin{centering}
\includegraphics[width=3in]{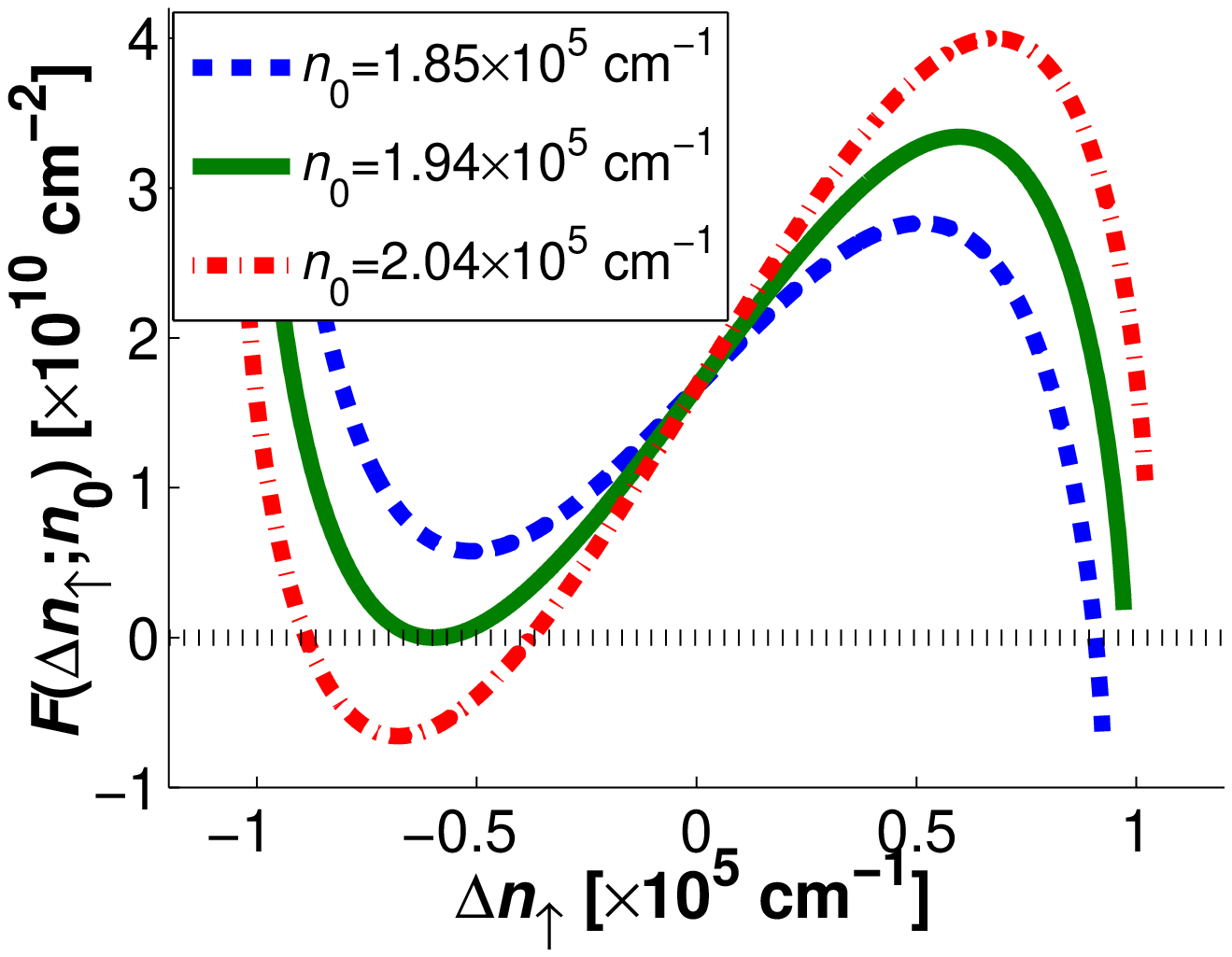}
\par\end{centering}

\protect\caption{\label{fig:S2}$F\left(\Delta n_{\uparrow};n_{0}\right)$ vs. $\Delta n_{\uparrow}$
for different $n_{0}$, with $B_{x}=\unit[1]{T}$ and $\hbar\omega_{y}=\hbar\omega_{z}=\unit[2]{meV}$.}
\end{figure}

\end{document}